\documentclass[aps,prb,twocolumn,showpacs,preprintnumbers,superscriptaddress,amsmath,amssymb,10pt]{revtex4-1}

\usepackage[T1]{fontenc}
\usepackage{mathcomp}
 
\usepackage{rotating}

\usepackage[justification=raggedright,singlelinecheck=false]{caption}

\begin{document}

\title{Newly observed first-order resonant Raman modes in few-layer MoS$_2$}

\author{Nils Scheuschner}\email{nils.scheuschner@tu-berlin.de}
\affiliation{Institut für Festkörperphysik, Technische Universität Berlin, Hardenbergstr. 36, 10623 Berlin, Germany}
\author{Roland Gillen}
\affiliation{Institut für Festkörperphysik, Technische Universität Berlin, Hardenbergstr. 36, 10623 Berlin, Germany}
\author{Matthias Staiger}
\affiliation{Institut für Festkörperphysik, Technische Universität Berlin, Hardenbergstr. 36, 10623 Berlin, Germany}
\author{Janina Maultzsch}
\affiliation{Institut für Festkörperphysik, Technische Universität Berlin, Hardenbergstr. 36, 10623 Berlin, Germany}

\date{\today}

\begin{abstract}
We report two new first-order Raman modes in the spectra of few-layer MoS$_2$ at 286~cm$^{-1}$ and  471~cm$^{-1}$ for excitation energies above 2.4~eV. These modes appear only in few-layer MoS$_2$; therefore their absence provides an easy and accurate method to identify single-layer MoS$_2$. We show that these modes are related to phonons that are not observed in the single layer due to their symmetry. Each of these phonons leads to several nearly degenerate phonons in few-layer samples. The nearly degenerate phonons in few-layer materials belong to two different symmetry representations, showing opposite behavior under inversion or horizontal reflection. As a result, Raman active phonons exist in few-layer materials that have nearly the same frequency as the symmetry forbidden phonon of the single layer. We provide here a general treatment of this effect for all few-layer two-dimensional crystal structures with an inversion center or a mirror plane parallel to the layers. We show that always nearly degenerate phonon modes of different symmetry must occur and, as a result, similar pseudo-activation effects can be excepted.
\end{abstract}


\maketitle

\section{Introduction}

\begin{figure}
\begin{minipage}{0.5\textwidth}
\includegraphics*[width=1.0\textwidth]{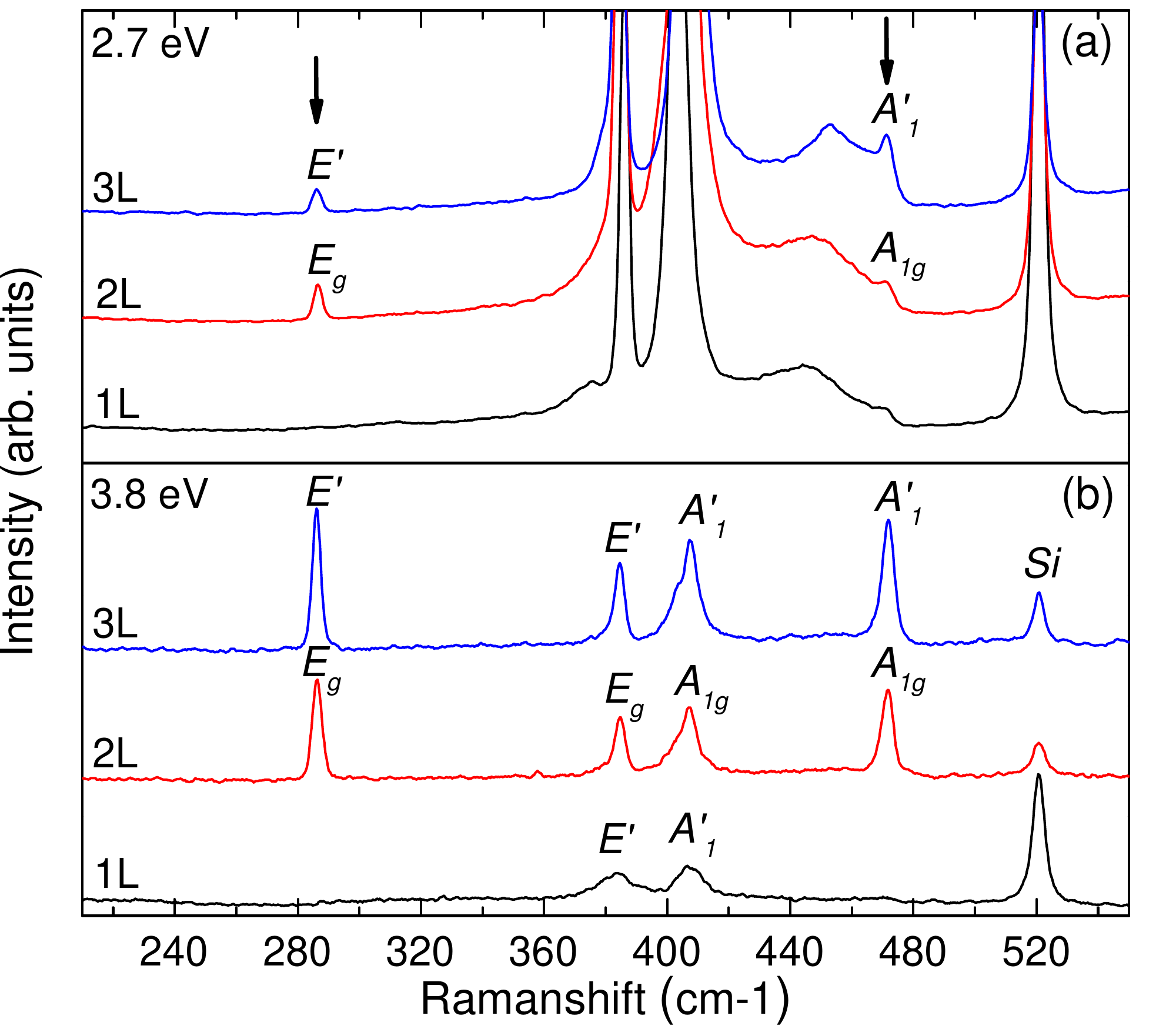}
\end{minipage}
\caption{\label{325} Raman spectra of single (1L), bi- (2L) and trilayer (3L) MoS$_2$ on Si/SiO$_2$ excited with (a) 2.7 eV and (b) 3.8 eV. The new modes are the $E'$ and $E_g$ modes around 286 cm$^{-1}$. Furthermore, an additional $A_{1g}$/$A'_1$ mode appears in few-layer MoS$_2$ 471 cm$^{-1}$, which is related to the $A''_2$ single-layer mode.}
\end{figure}

Two-dimensional crystals have received a lot of attention recently as they can have novel physical, chemical and mechanical properties not found in their bulk counterparts.\cite{Nicolosi2013,Geim2013} Besides graphene, layered transition metal dichalcogenides and especially molybdenum disulfide (MoS$_2$) show great potential for novel nanoelectronic and optoelectronic devices.\cite{Radisavljevic2011,Lembke2012,Buscema2013,Bertolazzi2011a,Ochedowski2014}
For characterizing layered materials, Raman and photoluminescence spectroscopy have been established as viable tools\cite{Mak2010,Tonndorf2013,Livneh2010,Scheuschner2014,Conley2013a,He2013,Chakraborty2012,Molina-Sanchez2011,Lee2010c}. We report here two new resonant first-order Raman modes for few-layer MoS$_2$. These modes appear in the Raman spectra for excitation energies above $2.4$~eV. At $2.7$~eV excitation energy, they are distinctly above the noise level; their intensity is comparable to second-order Raman peaks [see Fig.\ref{325}(a)]. At UV excitation, their intensity becomes even larger than the intensity of the typically examined $A_1'$/$A_{1g}$ and $E'$/$E_g$ modes [see Fig.\ref{325}(b)]. Therefore they can potentially be used as a quick and reliable way to discriminate between single- and few-layer MoS$_2$. 
Recently, similar Raman modes, which are either Raman inactive or forbidden in backscattering geometry in the bulk material, were reported in few-layer WSe$_2$, TaSe$_2$, and MoTe$_2$, however a systematic derivation of all vibrational modes in few-layer crystals, including their symmetry and displacement patterns is still lacking.\cite{Yamamoto2014,Terrones2014,Luo2014,Hajiyev2013} In this paper we present a detailed derivation of these modes using group theory. As the results follow from symmetry considerations alone, such Raman modes are not a unique feature of transition metal dichalcogenides, but are generally expected in all $N$-layer systems with an inversion symmetry center or a horizontal mirror plane.

All of the new Raman modes have in common that they are related to single-layer vibrations with the same frequency which are Raman inactive or are not observable in the specific scattering geometry.
Two effects are responsible for their appearance in the Raman spectra of few layer systems: 
$(i)$  The point group of the few-layer material changes compared to the single layer and depends on the stacking order and layer number. Therefore, also the representations of the vibration patterns in the few-layer system and their corresponding Raman tensors can change. 
$(ii)$ Each normal mode of the single-layer system leads to $N$ normal modes in the $N$-layer system. These modes are nearly degenerate in frequency in case of weak interlayer coupling. For a system with an inversion center or a horizontal mirror plane parallel to the layers, they belong to two different representations of the few-layer symmetry group, showing opposite character under inversion or horizontal reflection. For instance, a single-layer normal mode which is odd ($ungerade$) under inversion or horizontal reflection leads to both, even ($gerade$) and odd normal modes in the few-layer system. The even modes can now be Raman active, thus we observe a pseudo-activation of the inactive single-layer mode. The newly observed modes are thus not Raman inactive modes but regular Raman active modes of the few-layer material with nearly the same frequency as those forbidden in the single layer. 

To derive the vibration patterns of the few-layer normal modes, we discuss the applicability of a simple linear chain model combined with symmetry considerations and compare it to DFT calculations. The splitting of the single-layer modes into two different symmetries in the few-layer system will affect, besides Raman scattering, also all other physical processes where the symmetry of the normal modes is critical, such as electron-phonon scattering. Furthermore it is reasonable to assume to find a correspondingly splitting into nearly degenerate states of different symmetry also for the electronic wave functions.

\section{Vibrational properties and Raman scattering selection rules of molybdenum disulfide}

We will first discuss the vibrational properties and Raman selection rules for single-, bilayer and bulk ($2H$) MoS$_2$, as well as the Raman activation of the two new Raman modes for the bilayer, before we treat the case of a general $N$-layer system.
\subsection{Bulk molybdenum disulfide}
Bulk MoS$_2$ is a layered crystal, it consists of covalently bound \emph{single layers} of MoS$_2$, each formed by a layer of molybdenum atoms covalently bound to two layers of sulfur atoms. The individual MoS$_2$ layers in the bulk crystal are bound by van-der-Waals interaction. The structure of the most common polytype $2H$ consist of two parallel layers rotated relatively to each other by $\pi/6$, such that the sulfur atoms of one layer are directly below the molybdenum atom of the other layer. $2H$-MoS$_2$ belongs to the $D_{6h}$ symmetry group, with the $c$-axis perpendicular to the layers as the main rotational axis. The six atoms in the unit cell lead to 18 normal vibrations, which decompose at the $\Gamma$ point into the following irreducible representations:\cite{Verble1970,Ribeiro-Soares2014}
 
\begin{equation}
\Gamma _{2H}= A_{1g} \oplus 2A_{2u}\oplus 2B_{2g}\oplus B_{1u}\oplus E_{1g}\oplus 2E_{1u}\oplus 2E_{2g}\oplus E_{2u}
\end{equation}

The in-plane vibrations have $E$ symmetries and are twofold degenerate, while the out-of-plane vibrations have $A$ and $B$ symmetries. The $E_{2g}$, $E_{1g}$ and $A_{1g}$ symmetries correspond to Raman active modes. 
To describe the scattering geometries we use the notation $(\textit{\textbf{e}}_i,\textit{\textbf{e}}_s)$, where $\textit{\textbf{e}}_i$ ($\textit{\textbf{e}}_s$) is the polarization vector of the incoming (scattered) light. We define a cartesian coordinate system with the \textit{\textbf{x}} and \textit{\textbf{y}} vectors parallel to layer planes and the \textit{\textbf{z}} vector parallel to the \textit{c}-axis.
The $A_{1g}$ mode can be observed in (\textit{\textbf{x}},\textit{\textbf{x}}), (\textit{\textbf{y}},\textit{\textbf{y}}) and (\textit{\textbf{z}},\textit{\textbf{z}}) scattering geometries, while the $E_{2g}$ mode can be observed in (\textit{\textbf{x}},\textit{\textbf{x}}), (\textit{\textbf{y}},\textit{\textbf{y}}) and (\textit{\textbf{x}},\textit{\textbf{y}}) scattering geometries. In contrast, the $E_{1g}$ mode requires a scattering geometry involving a \textit{\textbf{z}}-component, \textit{i.e.} (\textit{\textbf{x}},\textit{\textbf{z}}), (\textit{\textbf{y}},\textit{\textbf{z}}) and (\textit{\textbf{z}},\textit{\textbf{z}}).\cite{CardonaII} All first-order Raman modes in bulk have been determined experimentally by Raman spectroscopy:\cite{Wieting1971,Chen1974} $\omega(E_{2g}^2)=32~$cm$^{-1}$, $\omega(E_{1g}^1)=287$~cm$^{-1}$, $\omega(E_{2g}^1)=383$~cm$^{-1}$ and  $\omega(A_{1g}^1)=409 $~cm$^{-1}$. Furthermore, the two IR-active optical phonons were measured by reflection spectroscopy:\cite{Wieting1971} $\omega(E_{1u})=384$~cm$^{-1}$ and $\omega(A_{2u})=470$~cm$^{-1}$. The $E_{2g}^1$ and $E_{1u}$ modes are nearly degenerate due to the low interaction between the layers.\cite{Verble1970}

\subsection{Single-layer molybdenum disulfide}

Single-layer molybdenum disulfide shows $D_{3h}$ symmetry. The unit cell consists of one metal and two sulfur atoms.
The representation of the nine normal modes at the $\Gamma$ point for the single layer can be decomposed into the following irreducible representations:\cite{Zhao2013}

\begin{equation}
\Gamma _{1L}^{\mathrm{MoS_2}} = A'_1 \oplus E'' \oplus  2 A''_2 \oplus 2 E'
\end{equation}

\begin{figure*}
\begin{minipage}{1.0\textwidth}
\includegraphics*[width=1.0\textwidth]{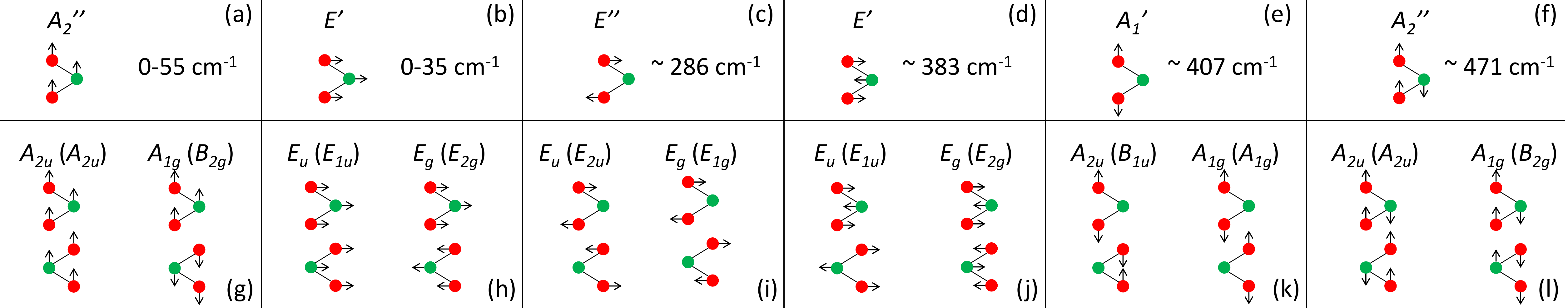}
\end{minipage}
\caption{\label{bilayer} (a)-(f) Normal mode vibration patterns and symmetries of single-layer MoS$_2$ (upper panels). The frequencies denote the typical range found in few-layer MoS$_2$. (g)-(l) The corresponding in-phase and anti-phase combination normal modes for bilayer (bulk) MoS$_2$ (lower panels).}
\end{figure*}

Figure \ref{bilayer} (a)-(f) shows the vibration patterns of the normal modes.
Due to the fact that the layers in the bulk are only weakly coupled, we find for all normal modes of the single layer two phonon modes of the bulk; Davydov pairs with only slightly modified frequencies and same displacement pattern (one with zero phase shift between the layers and one with a phase shift by $\pi$, see below).\cite{Lee2010c,Molina-Sanchez2011} The $A'_1$, $E'$ and $E''$ symmetries correspond to Raman active modes. The scattering geometries of these modes are equal to the scattering geometries of the corresponding Raman active modes $A_{1g}$, $E_{2g}$ and $E_{1g}$ in the bulk. The $E''$ mode requires therefore a scattering geometry with a \textit{\textbf{z}}-component to be observable. However as single-layer MoS$_2$ is a two dimensional crystal, it is difficult to realize such a scattering geometry experimentally in backscattering. Due to their similarity, it has become customary to associate the the $A'_1$ and $E'$ optical phonon modes of the single layer with the $A_{1g}$ and $E_{2g}$ modes of the bulk.

\subsection{Bilayer molybdenum disulfide}
\begin{figure}
\begin{minipage}{0.5\textwidth}
\includegraphics*[width=1.0\textwidth]{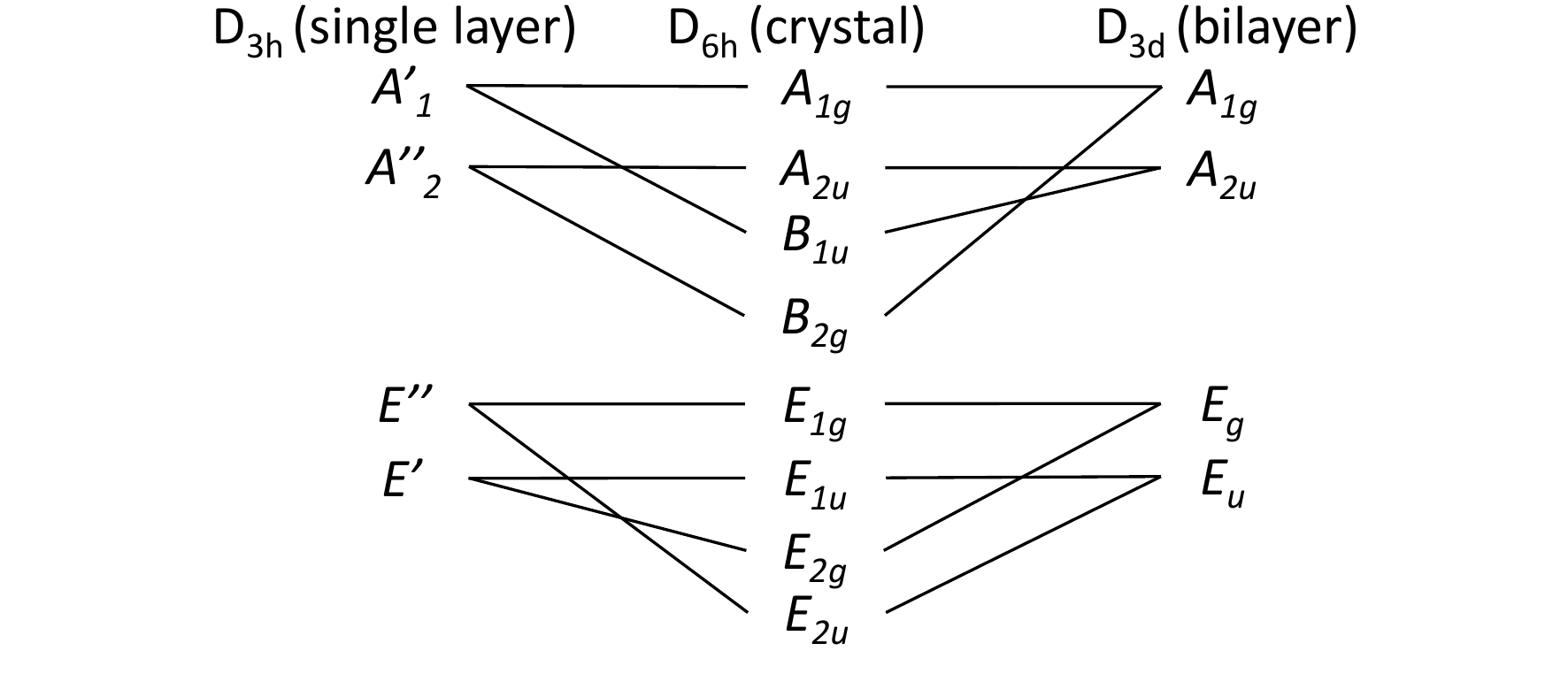}
\end{minipage}
\caption{\label{core} Correlation diagram of the symmetries of the normal modes for bulk ($2H$), single- and bilayer MoS$_2$.}
\end{figure}

The unit cell of bilayer molybdenum disulfide is the same as the one of $2H$ stacked bulk MoS$_2$. As a result, the vibration patterns of the bulk unit cell and the bilayer are equal [see Fig. \ref{bilayer} (g)-(l)]. However, the point groups are different, as the main rotation axis in bilayer is not six-fold like in bulk, but three-fold. The six-fold rotation symmetry of bulk MoS$_2$ results from a screw axis transformation, including the translation along the $c$-axis. As this transformation is not possible for a finite number of layers, bilayer MoS$_2$ belongs to the $D_{3d}$ point group. Due to the reduced symmetry, there are only four different representations in the bilayer instead of eight for the bulk. Figure \ref{core} shows the correlation of the symmetries of the normal modes between bulk, single- and bilayer MoS$_2$. 
The 18 normal vibrations decompose at the $\Gamma$ point into the following irreducible representations:\cite{Zhao2013}

\begin{equation}
\Gamma _{2L}^{\mathrm{MoS_2}} = 3 A_{1g} \oplus 3A_{2u}  \oplus  3E_g \oplus 3E_u
\end{equation}

Alternatively, the normal modes of the bilayer can be easily constructed from the single-layer normal modes: For each single-layer normal mode there are two normal modes of the bilayer obtained by combining the single-layer vibration in-phase and anti-phase (phase shift by $\pi$), as shown in Fig. \ref{bilayer}. The in-phase and anti-phase combinations always show opposite transformation behavior under spatial inversion and thus belong to two different symmetry representations, \textit{i.e.} $gerade$ and $ungerade$. On the other hand, they have nearly the same phonon energy in case of weak interlayer coupling. This leads to the generation of a Raman active mode in the bilayer from a previously Raman inactive single layer mode: From the single-layer $A''_2$ mode an IR-active $A_{2u}$ mode and a Raman active $A_{1g}$ mode are constructed in the bilayer [Fig. \ref{bilayer}~(l)]. This has been experimentally observed for WSe$_2$, TaSe$_2$, and MoTe$_2$ in Ref. \onlinecite{Yamamoto2014,Terrones2014,Luo2014,Hajiyev2013}. As the $A_{2u}$ bulk phonon in MoS$_2$ has been measured by IR spectroscopy and calculations of the single-layer $A''_2$ phonon energy correspond very well to one of the newly observed modes in the bilayer (Fig. \ref{325}), we assign this new mode at 471~cm$^{-1}$ to the bilayer $A_{1g}$ vibration. In the bulk material the corresponding vibration pattern has $B_{2g}$ symmetry and is therefore Raman inactive. The $A_{2u}$ phonon of the bilayer is in bulk still described by an $A_{2u}$ representation and remains IR active.

For the $E''$ single-layer mode we observe a similar effect regarding the Raman scattering geometry. The $E''$ mode is Raman active in the single-layer, however it can not be observed in backscattering geometry. In bilayer with $D_{3d}$ symmetry, we find now an $E_g$ and an $E_u$ mode originating from the single-layer $E''$ mode [Fig. \ref{bilayer}~(i)]. The $E_g$ mode of the bilayer is -- like the $E''$ mode of the single layer -- Raman active, but in contrast also observable in backscattering. Again, we find a good agreement between the newly observed mode at 286~cm$^{-1}$ (Fig.\ref{325}) and calculations, as well as with Raman measurements of the bulk $E_{1g}$ phonon. Therefore, we assign this new mode to the bilayer $E_g$ symmetry.

\section{Normal vibration modes and their symmetry representations in an $N$-layer system}

The concept of combining the single-layer normal modes in-phase and anti-phase can be extended to a $N$-layer system with an inversion center or a mirror plane parallel to the layers. For the sake of comprehensibility, we neglect in the following the degeneracy of the $E$-type normal modes, as it will not affect the results. Generally, the number of normal modes is proportional to the number $N$ of layers, as the number of atoms in the unit cell of a few-layer material increases linearly with $N$. For normal modes all atoms move periodically with the same frequency. Thus the few-layer normal modes can be constructed only by a superposition of identical single-layer vibration patterns with possibly different amplitudes in each layer (assuming mode mixing effects to be negligible, due to weak interlayer interaction). These amplitudes form a vector ${\textit{\textbf{n}}} \in\Re^N$, with the amplitude of the single-layer normal mode in the $j$th-layer as the $j$th-component $n_j$ of the vector $\textit{\textbf{n}}$. The normal modes of the few-layer system, originating from a given single-layer normal mode, are now described by a set of vectors $\textit{\textbf{n}}^{(i)} \in\Re^N, i= \left\{1,..,N \right\}$, which form a basis for the $\Re^N$, as the normal modes are required to be linearly independent. To find the symmetries of the few-layer normal modes and the number of modes with a certain symmetry, it is not necessary to determine the $\textit{\textbf{n}}^{(i)}$ (which requires to find and solve the dynamical matrix); instead it is sufficient to consider the general behavior of the $\textit{\textbf{n}}^{(i)}$ under the symmetry transformation of the few-layer point group, as shown in the following.

\begin{figure}
\begin{minipage}{0.5\textwidth}
\includegraphics*[width=1.0\textwidth]{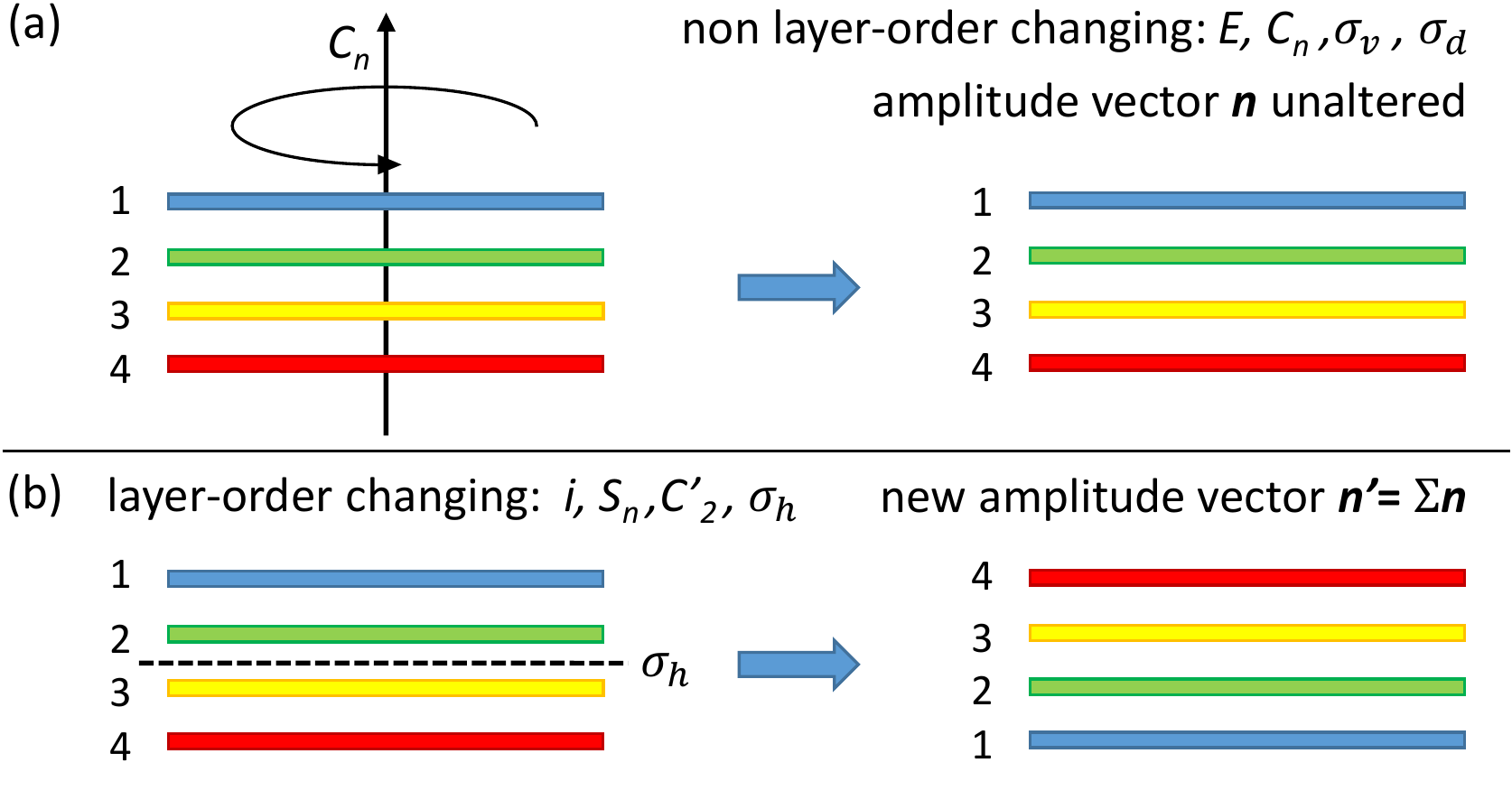}
\end{minipage}
\caption{\label{layerorder} Illustration of the effect of (a) the non layer-order changing operations and (b) the layer-order changing operations on a 4-layer structure.}
\end{figure}

Layered materials can not belong to the polyhedreal point groups, \textit{i.e.}, $T,T_d,T_h,O$, and $O_h$ in Schoenflies notation, as those groups contain rotations, whose rotation axes are not parallel or perpendicular to the layer planes and would therefore not transform them into themselves. All layered materials must therefore belong to one of the remaining 27 point groups. The symmetry transformations found in these groups can be divided into two sets, depending on their effect on the layer order in a few-layer structure: one set of symmetry operations that do not change the order of the layers, \textit{i.e.}, $\left\{E,C_n,\sigma _v,\sigma _d\right\}$ in Schoenflies notation, and one set with the symmetry operations that do change the layer order $\left\{i,S_n,C'_2,\sigma_h\right\}$ (see. Fig. \ref{layerorder}). 
Among the 27 point groups there are 9 groups that contain only non-layer-order changing operations, \textit{i.e.}, ${C_1, C_n}$ and $ C_{nv}$ with $n=2,3,4,6$. Furthermore, there are 12 groups that have an inversion center or a mirror plane parallel to the layers, thus being the point groups relevant for the few-layer systems considered here. These 12 groups can always be written as a direct product of one of the 9 groups without layer-order changing operations with $C_i$ or $C_{1h}$ (see Tab. \ref{tab:multi}).
\begin{table}
\caption{\label{tab:multi} Multiplication table (direct product) of the 9 point groups that contain only non layer-order changing operations with $C_i$ and $C_{1h}$. All 12 point groups that contain an inversion center or horizontal reflection plane symmetry can be written as such a product.}
\begin{ruledtabular}
\begin{tabular}{llllllllll}
 & $C_{1}$ &  $C_{2}$  &  $C_{3}$ &  $C_{4}$ &  $C_{6}$ &  $C_{2v}$  &  $C_{3v}$ &  $C_{4v}$ &  $C_{6v}$\\
	\hline
$C_{i}$    & $C_{i}$ &  $C_{2h}$   &  $S_{6}$ &   $C_{4h}$ &  $C_{6h}$ &  $D_{2h}$  &  $D_{3d}$ &  $D_{4h}$ &  $D_{6h}$\\
$C_{1h}$   & $C_{1h}$ &  $C_{2h}$  &  $C_{3h}$ &  $C_{4h}$ &  $C_{6h}$ &  $D_{2h}$  &  $D_{3h}$ &  $D_{4h}$ &  $D_{6h}$\\

\end{tabular}
\end{ruledtabular}
\end{table}
As a result the few-layer point groups have always a uniquely defined subgroup that contains only non-layer-order changing operations. In the following we will denote this group as the non-layer-order changing subgroup.

Obviously, under all symmetry operations that do not alter the order of the layers, the $N$-layer normal modes will transform like the corresponding single-layer vibration pattern.
For the layer-order changing operations, we can focus on the $i$ and $\sigma_h$ operations, as all further layer-order-changing operations can be written as a product of an element of the non-layer-order-changing subgroup with $i$ or $\sigma_h$ (\textit{e.g.} $C'_2=\sigma_v\sigma_h$). As a result, the behavior under inversion or horizontal reflection will determine the behavior of all other layer-order changing operations.

In the following, we will focus on $N$-layer systems with inversion symmetry. All results can be directly transferred to systems with horizontal mirror symmetry simply by interchanging $i$ and $\sigma_h$. We define now an operator $\Sigma$ that inverts the order of the components of the vector it is applied to: 

\begin{equation}
 \Sigma \textit{\textbf{v}}   =\Sigma \left(v_1,... ,v_{j-1},v_j\right)= \left(v_j,v_{j-1},... ,v_1\right) 
\end{equation}

\begin{figure}
\begin{minipage}{0.5\textwidth}
\includegraphics*[width=1.0\textwidth]{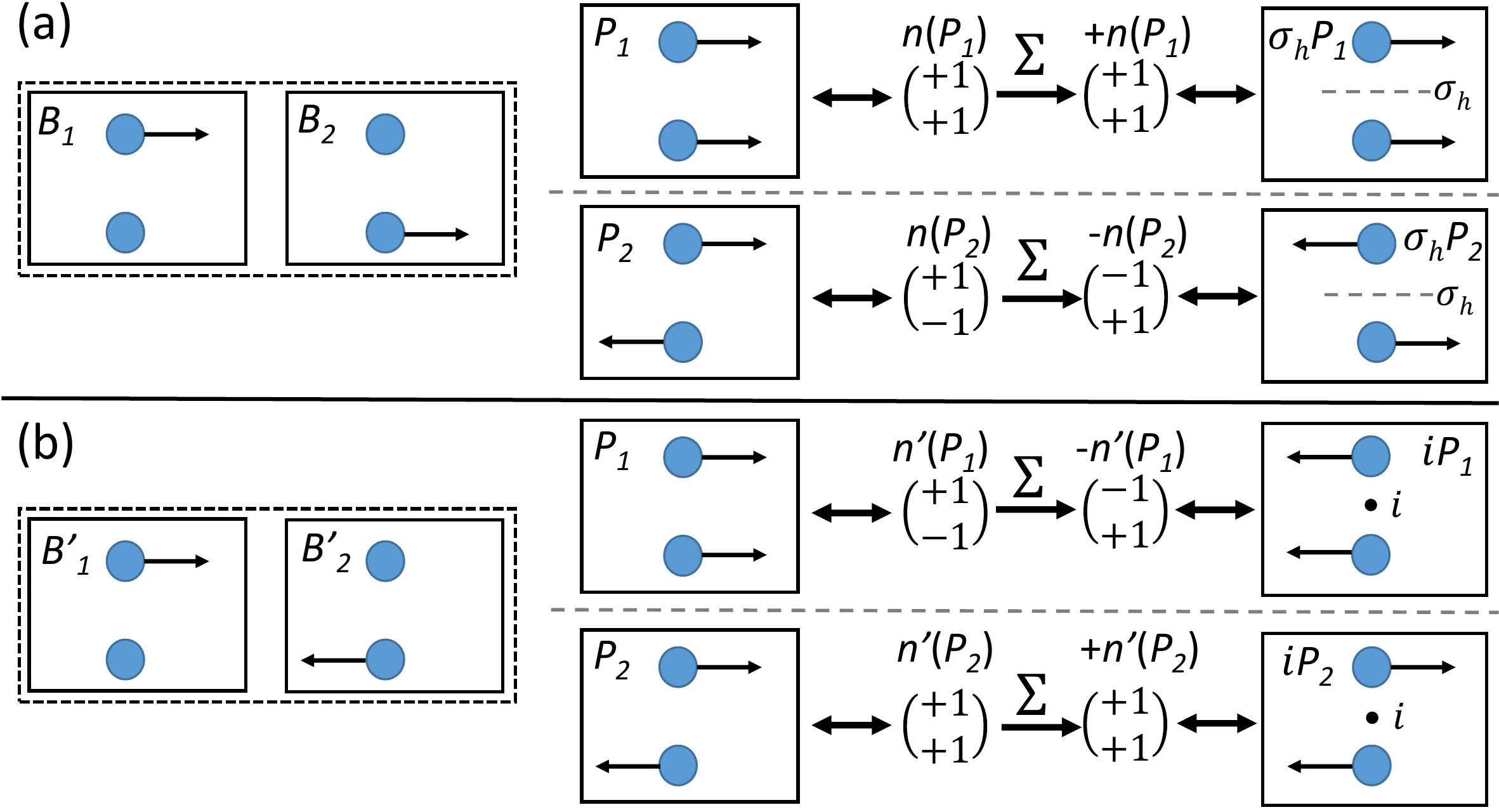}
\end{minipage}
\caption{\label{basis} Schematic illustration of the effect of the choice of the basis displacement patterns. Two different bases are defined by the displacement patterns $B_1$ and $B_2$ for (a) and $B'_1$ and $B'_2$ for (b). In each basis the same two displacement patterns ($P_1$ and $P_2$) are constructed, due to the different bases the corresponding vectors $n$ and $n'$ are unequal. Applying $\Sigma$ to the vectors $n$ and $n'$ results for (a) in displacement patterns that are reflected along a horizontal mirror plane, or are inverted at an inversion center for (b).}
\end{figure}

Inverting the few-layer vibration pattern leads effectively to an inversion of the order of the amplitudes -- just like the effect of $\Sigma$. However the inversion can also change the amplitude signs depending on the definition of the basis displacement patterns that are used to construct the displacement patterns from a vector $\textit{\textbf{n}}^{(i)}$. This effect is illustrated for a bilayer system in Fig. \ref{basis}. 

Nevertheless, for an even layer number, it is always possible to find a basis where applying $\Sigma$ to a vector will result in a vector describing the inverted displacement pattern, \textit{e.g.} like in Fig. \ref{basis}(b), by choosing all pairs of basis vectors that are transformed onto each other under inversion accordingly.
In this case the vectors $\textit{\textbf{n}}^{(i)}$, will be eigenvectors of $\Sigma$ with eigenvalue $\kappa$ that can only become $\pm 1$. The sign of $\kappa$ decides now how the corresponding few-layer vibration pattern transforms under inversion. 
Obviously, for even layer numbers, the few-layer normal mode will be even ($gerade$) for positive $\kappa$ and odd ($ungerade$) for negative. 

For odd layer numbers we have to consider that the central layer will be transformed onto itself and not onto another layer. As a result the few-layer vibration patterns can have only the same transformation behavior under inversion as the single layer vibration pattern, when the central layer has an displacement amplitude unequal zero.
Furthermore we have to use $-\Sigma$ instead of $\Sigma$ to find the vector that describes the inverted displacement pattern when the corresponding single-layer vibration is odd. On the other hand, a few layer vibration, where the central layer shows non-zero displacement can be only described by an vector $\textit{\textbf{n}}^{(i)}$ with $\kappa=+1$, as the amplitude of the central layer will be left unaltered by $\Sigma$. All remaining basis vectors can be chosen such that their behavior under inversion matches the behavior of the central layer. As a result using such a basis we find that for $\kappa=+1$, the few-layer vibration will now behave exactly as the single-layer vibration under inversion, whereas it shows opposite behavior for $\kappa=-1$. 

We now turn to the question how many of the $N$ normal modes created by a given single-layer normal mode have $\kappa=+1$ or $\kappa=-1$. 
This question can be answered independently of the inter- and intralayer coupling, as the eigenvectors of $\Sigma$ divide $\Re^N$ into two orthogonal subspaces depending on the eigenvalue.\cite{Note1}
As the vectors $\textbf{n}^{(i)}$ form a basis for their respective subspaces and simultaneously form a basis for $\Re^N$, the absolute number of the normal modes with the same eigenvalue can not change, regardless the amount of coupling. Therefore we always find $\left\lceil \frac{N}{2} \right\rceil$ (ceiling function of $\frac{N}{2}$) normal vibrations with $\kappa=+1$ and $\left\lfloor \frac{N}{2} \right\rfloor$ (floor function of $\frac{N}{2}$) with $\kappa=-1$.

\begin{figure}
\begin{minipage}{0.5\textwidth}
\includegraphics*[width=1.0\textwidth]{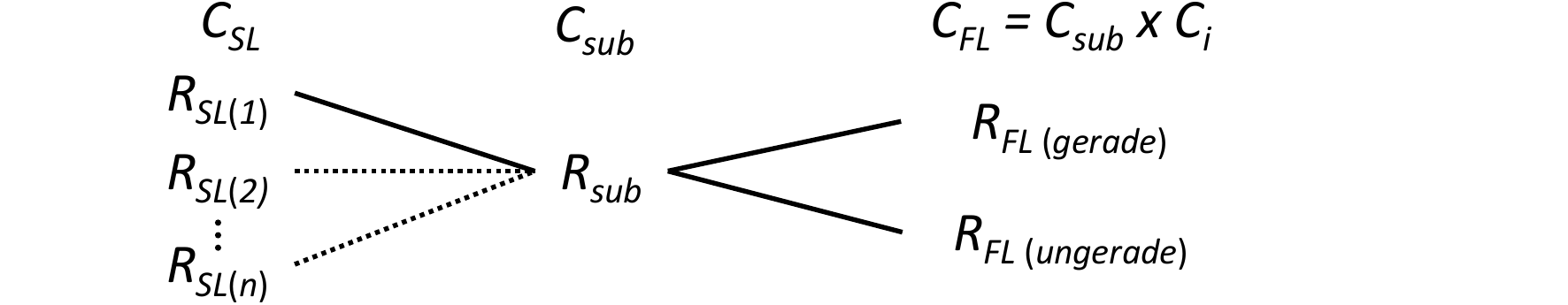}
\end{minipage}
\caption{\label{reps} Correlation diagram of the normal modes in the single-layer point group ($C_{SL}$), the few-layer point group ($C_{FL}$), and the non-layer-order changing subgroup ($C_{sub}$). Several normal modes of the single layer can reduce to the same representation of $C_{sub}$, indicated by the dotted lines. In contrast always two normal modes of the few-layer system, one $gerade$ and one $ungerade$, reduce to the same representation of $C_{FL}$.}
\end{figure}

The symmetry of the few-layer normal modes can now be easily found. As the few-layer normal mode transforms under the non-layer-order changing transformations exactly as the single-layer normal mode used for the construction of the few-layer normal mode, we determine the reduced representation of the single-layer normal mode in the non-layer-order changing subgroup of the few-layer point group. Starting from this representation, the few-layer point group, which is the product group of the the non-layer-order changing subgroup with $C_i$, contains two representations, one $gerade$ and one $ungerade$. These are the possible symmetries of the few-layer normal modes, see Fig. \ref{reps}.

\subsection{Application to few-layer molybdenum disulfide}

The point group of few-layer MoS$_2$ with $2H$-like stacking depends on the layer number, see above and Ref. \onlinecite{Ribeiro-Soares2014}. Single-layer MoS$_2$ has a horizontal mirror plane; in odd layer numbers this symmetry is preserved, as a result they belong to the $D_{3h}$ point group. Structures with an even layer number have an inversion center instead of the horizontal mirror plane; they belong to the $D_{3d}$ point group. Both groups have the same non-layer-order changing subgroup $C_{3v}$ (see Tab. \ref{tab:multi}).

At the $\Gamma$ point of single-layer MoS$_2$ we find six different phonons which transform according to the $A'_1$, $A''_2$, $E'$ and $E''$ representations of the $D_{3h}$ point group. The reduced representation of these phonons in $C_{3v}$ are $E$ for $E'$ and $E''$ and $A_1$ for $A'_1$ and $A''_2$. To determine the possible representations of the phonons in a few-layer structure with even layer number, we search the representations in $D_{3d}$ that reduce to $E$ and $A_1$ in $C_{3v}$. These are $E_g$ and $E_u$ for $E$, and $A_{1g}$ and $A_{2u}$ for $A_1$. 
For odd layer numbers, using the same approach, we search the representations in $D_{3h}$ that reduce to $E$ and $A_1$ in $C_{3v}$. These are $E'$ and $E''$ for $E$, and $A'_1$ and $A''_2$ for $A_1$. The results of the previous section, concerning the number of normal modes with $\kappa=+1$ and $\kappa=-1$, allow us now to determine number of the phonons with a specific symmetry in the few-layer crystals that are constructed from each single-layer normal mode (see Table \ref{tab:apli}). Due to interlayer coupling the degeneracy of the few-layer phonons will be lifted. However, for the \textit{N}-layer normal modes originating from the optical single-layer phonons this effect is quite small. For the $N$-layer normal modes originating from the acoustic phonons, on the other hand, the interlayer coupling leads for all but one modes to a non-zero phonon energy. Due to their vibration patterns, these optical modes are denoted as shear and layer breathing-like modes and have been extensively reported in the literature.\cite{Plechinger2012,Zeng2012a,Zhang2013,Zhao2013}

From the $E''$ and $A''_2$ phonons of the single-layer the general analysis discussed above naturally predicts normal modes with $E{_g}/E'$ symmetry at 286~cm$^{-1}$ (see Fig. \ref{bilayer}~(i) and Fig.~\ref{constr}) and $A_{1g}/A'_1$ symmetry at 471~cm$^{-1}$ [see Fig. \ref{bilayer}~(l)] in few-layer MoS$_2$. These phonons are Raman active and explain the the newly observed Raman modes in Fig. \ref{325}. From a symmetry point-of-view, these modes always appear in few-layer structures with $N>1$.  Many of the transition metal dichalcogenides have a structure and stacking order like $2H$-MoS$_2$. While the phonon energies will certainly change, their symmetries and therefore also their Raman selection rules follow the same structure as discussed above. Thus similar modes to the newly observed few-layer modes are expected and have partially been observed in some isostructural materials.\cite{Yamamoto2014,Terrones2014,Luo2014,Hajiyev2013}

\begin{table}
\caption{\label{tab:apli} Number of $\Gamma$-point phonon modes and their representation of the point group of single-, few-layer and bulk MoS$_2$. $\omega_0$ gives the typical range of the phonon frequencies. Bold letters denote the symmetries of Raman-active phonons which are allowed in backscattering geometry with the incoming light propagating perpendicular to the layer plane. Stars ($\star$) denote the newly observed Raman modes.}
\begin{ruledtabular}
\begin{tabular}{llrlll}
MoS$_2$  & $D_{3h}$ &   & $D_{3d}$ & $D_{3h}$ & $D_{6h}$ \\
$\omega_0$ (cm$^{-1}$) & $N$=1 & $\kappa$ &$N$ even & $N$ odd  & bulk \\
\hline
0-55     & $A''_2 $& +1 & $\frac{N}{2}\cdot \mathbf{A_{1g}}$  & $\frac{N+1}{2}\cdot         A''_2$ & $B_{2g}$\\  
          &    -   & -1 & $\frac{N}{2}\cdot A_{2u}$           & $\frac{N-1}{2}\cdot \mathbf{A'_1}$ & $A_{2u}$\\ 
 
0-35 & $\mathbf{E'}$   & +1&  $\frac{N}{2}\cdot \mathbf{E_g}$  & $\frac{N+1}{2}\cdot \mathbf{E'} $& $\mathbf{E_{2g}}$\\
     & -      & -1&  $\frac{N}{2}\cdot E_u$           & $\frac{N-1}{2}\cdot E''$         & $E_{2u}$\\
285-287 & $E''$  & +1&$\frac{N}{2}\cdot \mathbf{E_g(\star)} $& $\frac{N+1}{2}\cdot E''      $ & $E_{1g}$\\
        & -      & -1&$\frac{N}{2}\cdot  E_u$           & $\frac{N-1}{2}\cdot \mathbf{E'(\star)}  $     & $E_{1u}$\\
382-385 & $\mathbf{E'}$   & +1&$\frac{N}{2}\cdot \mathbf{E_g}$& $\frac{N+1}{2}\cdot \mathbf{E'}$& $\mathbf{E_{2g}}$\\
        & -               & -1&$\frac{N}{2}\cdot E_u$         & $\frac{N-1}{2}\cdot E''        $ & $E_{2u}$\\
402-409  & $\mathbf{A'_1}$  & +1&$\frac{N}{2}\cdot \mathbf{A_{1g}}$ & $\frac{N+1}{2}\cdot \mathbf{A'_1}   $     &   $\mathbf{A_{1g}}$ \\
         & -                & -1&$\frac{N}{2}\cdot A_{2u}$          & $\frac{N-1}{2}\cdot  A''_2 $&  $B_{1u}$\\
470-471  & $A''_2$ & +1&$\frac{N}{2}\cdot \mathbf{A_{1g}(\star)}$  & $\frac{N+1}{2}\cdot A''_2$          & $B_{2g}$\\
         & -       & -1&$\frac{N}{2}\cdot A_{2u}$             & $\frac{N-1}{2}\cdot \mathbf{{A'_1}(\star)}$& $A_{2u}$\\
\end{tabular}
\end{ruledtabular}
\end{table}

\subsection{Application to $AB$-stacked graphene}

Monolayer graphene belongs to the $D_{6h}$ point group, and has therefore an inversion center and a horizontal reflection plane. In few-layer graphene (with $AB$ stacking), the six-fold rotation axis is lost; instead it shows a three-fold main rotation axis. Also the inversion center and horizontal reflection plane are not always preserved in the few-layers: Few-layer graphene with an even layer number has only an inversion center. With an odd number of layers it has only the horizontal reflection plane. Therefore, the point group is $D_{3d}$ for even layer numbers and $D_{3h}$ for odd layer numbers with $N>1$, like in MoS$_2$.\cite{Malard2009a} The subgroup that does not contain the layer-order-changing operations is in both cases $C_{3v}$. The monolayer has six $\Gamma$ point phonons of $A_{2u}$, $B_{2g}$, $E_{1u}$ and $E_{2g}$ symmetry. The reduced representation of the $A_{2u}$ and $B_{2g}$ modes in $C_{3v}$ is $A_{1}$; $E_{1u}$ and $E_{2g}$ reduce to $E$. This leads to the same possible symmetries for the few-layer graphene phonons as for few-layer MoS$_2$ with the same point group (see Tab. \ref{tab:apliG}). Again, the acoustic phonons of single-layer graphene lead to the optical shear and layer breathing like modes.\cite{Michel2008,Tan2012,Herziger2012,Lui2013} The $E_{2g}$ phonon, which is responsible for the $G$ mode of graphene, has Raman active counterparts of $E_g/E'$ symmetry in the few-layer structures. Notably, the $B_{2g}$ mode leads to normal modes of $A_{1g}/A'_1$ symmetry in few-layer graphene that may be observable in Raman spectroscopy from a symmetry point of view. However, to our best knowledge, no experimental observation of these modes has been reported so far. 

\begin{table}
\caption{\label{tab:apliG} Number of $\Gamma$-point phonon modes of single-, few-layer graphene and graphite and their representation of the point group. $\omega_0$ is the typical range of the phonon frequencies. Bold letters denote the symmetries of Raman-active phonons which are allowed in backscattering geometry with the incoming light propagating perpendicular to the layer plane.}
\begin{ruledtabular}
\begin{tabular}{llrllll}
Graphene  & $D_{6h}$ &   & $D_{3d}$ & $D_{3h}$ & $D_{6h}$ \\
$\omega_0$ (cm$^{-1}$) & $N$=1 & $\kappa$ &$N$ even & $N$ odd  ($N>1$) & bulk \\

\hline
0-127     & $A_{2u} $& +1 &$\frac{N}{2}\cdot \mathbf{A_{1g}}$  & $\frac{N+1}{2}\cdot A''_2$          & $B_{2g}$\\
         &    -      & -1 &$\frac{N}{2}\cdot A_{2u}$           & $\frac{N-1}{2}\cdot \mathbf{A'_1}$  & $A_{2u}$\\

0-42 & $E_{1u}$   & +1&$\frac{N}{2}\cdot \mathbf{E_g}$ & $\frac{N+1}{2}\cdot \mathbf{E'}$ & $\mathbf{E_{2g}}$\\
     & -          & -1&$\frac{N}{2}\cdot E_u$ & $\frac{N-1}{2}\cdot E''$& $E_{1u}$\\

868 & $B_{2g}$  & +1&$\frac{N}{2}\cdot \mathbf{A_{1g}}$& $\frac{N+1}{2}\cdot A''_2 $& $B_{2g}$\\
       & -      & -1&$\frac{N}{2}\cdot A_{2u}$& $\frac{N-1}{2}\cdot \mathbf{A'_1}  $& $A_{2u}$\\

1582 & $\mathbf{E_{2g}}$   & +1&$\frac{N}{2}\cdot \mathbf{E_g}$   & $\frac{N+1}{2}\cdot \mathbf{E'} $ & $\mathbf{E_{2g}}$\\
       & -                 & -1&$\frac{N}{2}\cdot E_u$   & $\frac{N-1}{2}\cdot E''$ & $E_{1u}$\\

\end{tabular}
\end{ruledtabular}
\end{table}

\subsection{Constructing the $N$-layer normal modes}

We will now show how to use a simple linear-chain model to construct the displacement patterns of the few-layer vibrational modes. 
The application of this model\cite{Luo1996} is well established for the low frequency modes in few-layer graphene and other layered materials.\cite{Tan2012,Zhang2013,Zhao2013} 
In this model each layer is represented by a classical harmonic oscillator which interacts only with it nearest neighbors directly. Such a system is similar to a resonator with open ends, whose solutions are standing waves.\cite{Gillen2009} For a finite number of oscillators, those standing waves form an envelope for the eigenvectors $\textit{\textbf{c}}^{(i)}$ of the solutions -- indexed by $i$ -- whose components $c^{(i)}_{j}$ describe the amplitude of the $j$-th layer. Furthermore the finite number of oscillators limits the minimum wavelength of the standing waves; as a result there are $N$ solutions for an $N$-layer system. 

The normal modes originating from the single-layer acoustic modes can be easily constructed from the vectors $\textit{\textbf{c}}^{(i)}$, as the for the single-layer acoustic modes all atoms show the same displacement.
Thus the product of the single-layer acoustic mode displacement vector with the vector component $c^{(i)}_{j}$ can be used directly as the displacement vector for all atoms in the $j$th-layer. 

\begin{figure}
\begin{minipage}{0.5\textwidth}
\includegraphics*[width=1.0\textwidth]{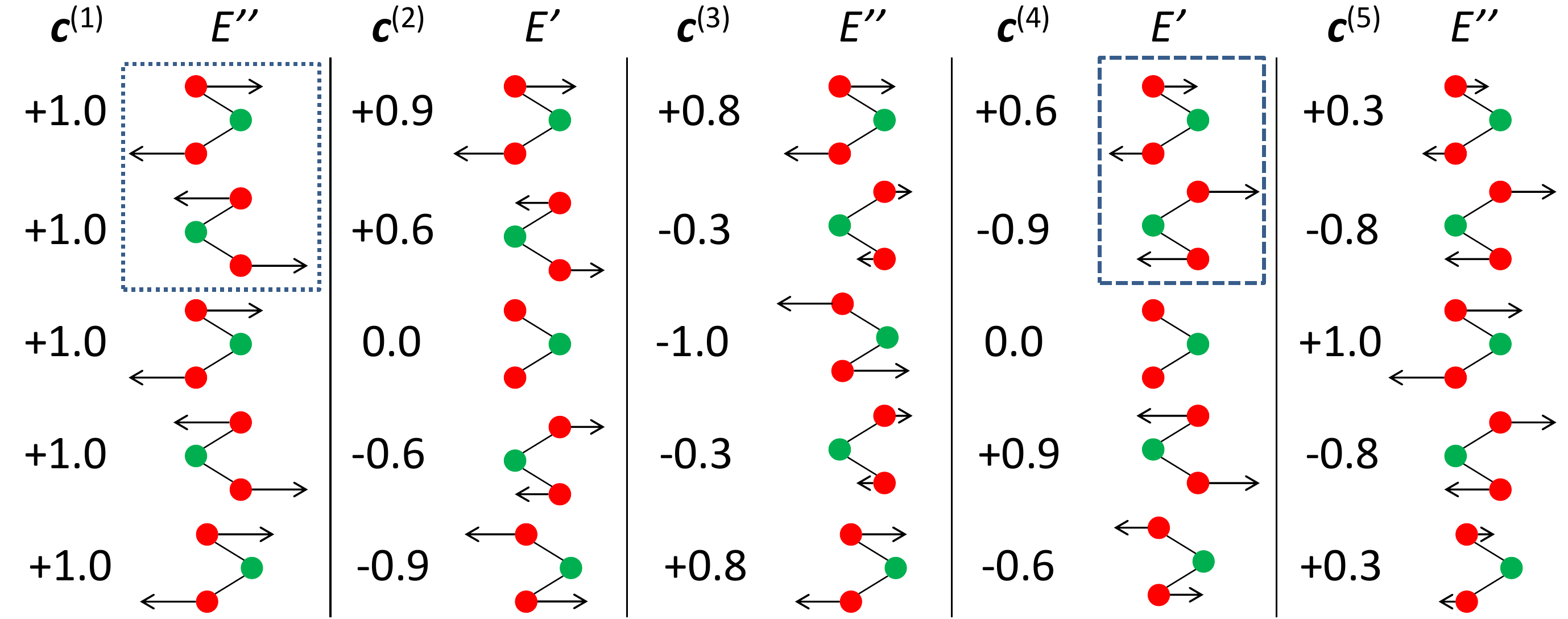}
\end{minipage}
\caption{\label{constr} The normal modes of five-layer MoS$_2$ originating from the optical $E''$ phonon of the single layer. They are constructed using the eigenvectors $\mathrm{\textbf{c}}^{(i)}$ of the solution of the linear chain model. The dotted (dashed) box shows the combination of single-layer vibrations leading to lower (higher) interlayer coupling energy. The vibration patterns calculated by DFT are the same despite small mode mixing effects with modes of the same symmetry.}
\end{figure}

For the few-layer normal modes generated from optical phonons, the individual atoms in each layer show different displacement vectors. Therefore it is necessary to consider which combination of the single layer vibration pattern leads to a higher or lower interlayer coupling energy for neighboring layers. This can be easily derived from a bilayer system: Starting from the top layer ($j=1$) we construct the vibration pattern of the few-layer system step by step, by applying the single-layer vibration pattern to the next layer such that it minimizes (maximizes) the interlayer coupling when the sign of the components $c^{(i)}_{j}$ of the neighboring layers are the same (opposite). In Figure \ref{constr} the construction of the normal modes originating from the $E''$ (286 cm$^{-1}$) single-layer phonon leading to the newly observed Raman modes is demonstrated for five-layer MoS$_2$. 
The symmetries of the few-layer systems can be easily attributed to the two possible representations by optical inspection, see above. As expected, we find two Raman active $E'$ modes and three inactive $E''$ modes. 

We have calculated the normal modes of the layered transition metal dichalcogenide WS$_2$ by density functional theory (DFT), a material where the layers are relatively strongly bound, at the $\Gamma$-point for up to five layers.\cite{Note2,Staiger2015}
Using the scalar product of the displacement vectors as a metric for the similarity of the normal modes, we find a perfect match for the low-frequency modes, as those modes originate from acoustic modes. For the higher-energy optical modes, we find a small mixing effect of modes with the same symmetry, which increases with the layer number. However, for five-layer WS$_2$, we still find an average similarity of 93.4\% of the linear-chain model normal modes compared to the DFT results. 

We expect all other layered materials to behave similarly; therefore, unless the displacement patterns are required with high accuracy, the linear-chain model might be used to find the vibration patterns of the high-energy normal modes.

\section{Resonance effects of Raman modes in MoS$_2$}

We prepared single and few-layer MoS$_2$ flakes by mechanical exfoliation on Si substrates with an SiO$_2$ layer of 50nm and 90nm thickness from natural MoS$_2$ (SPI supplies) and freestanding bilayer MoS$_2$ as described in Ref. \onlinecite{Scheuschner2014}. We determined the layer number by measuring the low-frequency Raman modes.\cite{Plechinger2012,Zhang2013,Zhao2013} Raman measurements were performed with a Horiba LabRAM HR with 458~nm, 532~nm and 830~nm diode lasers, a T64000 triple monochromator system with an Ar-ion laser and a HeCd-laser, and a Dilor XY800 spectrometer with a Kr-ion laser and a dye laser. The Raman measurements were performed at room temperature in a confocal setup using a 100x objective except for the UV measurements, where a 40x objective was used. For all measurements the laser power was below 0.5~mW to avoid heating effects.

\begin{figure}
\begin{minipage}{0.5\textwidth}
\includegraphics*[width=1.0\textwidth]{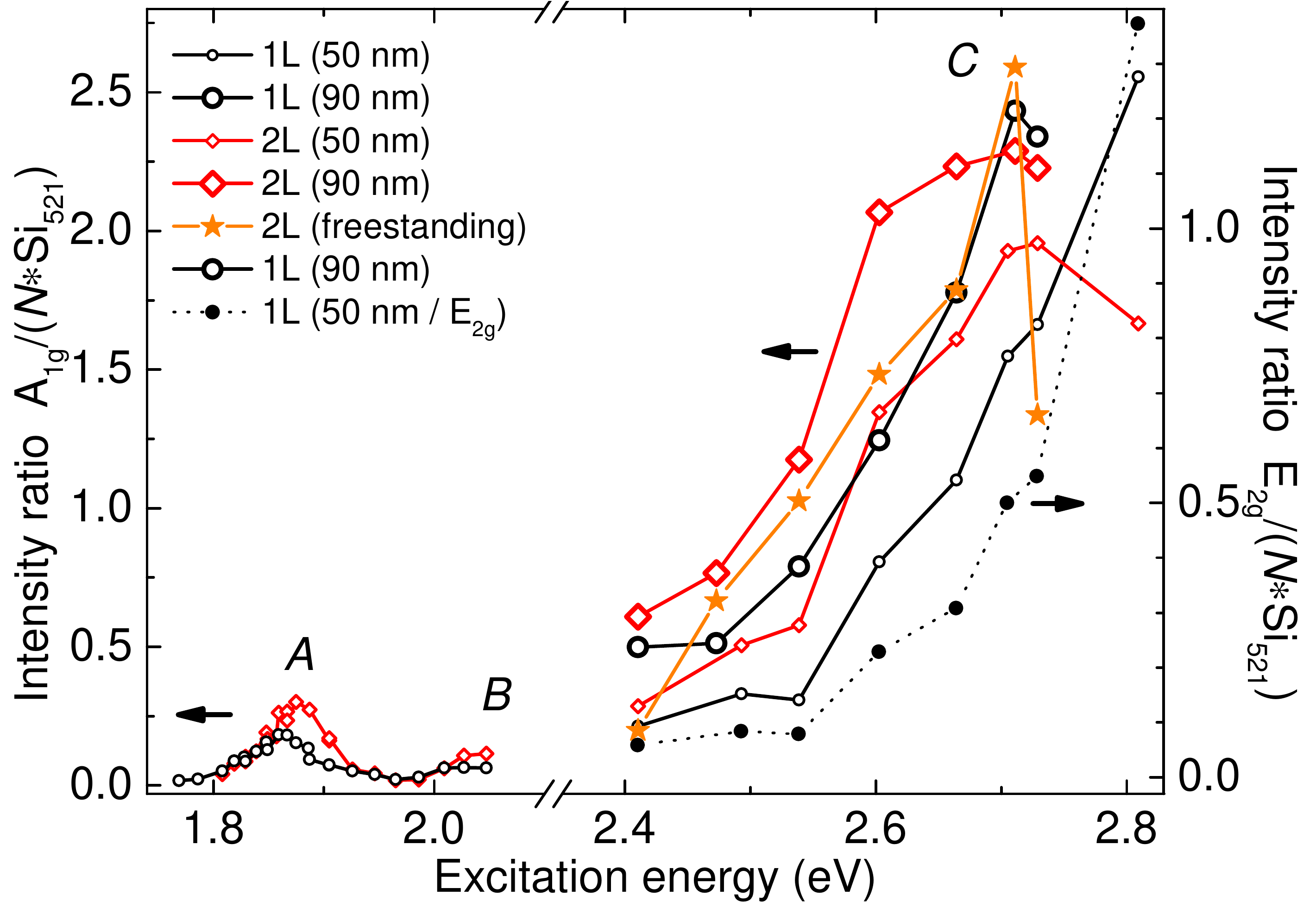}
\end{minipage}
\caption{\label{A1G_ratio} Intensity ratio of the normally observed Raman modes and the 521~cm$^{-1}$ mode of the silicon substrate, normalized to the number of MoS$_2$ layers. Hollow (filled) symbols indicate the $A_{1g} / A'_1$ $\approx$ 407~cm$^{-1}$ ($E_{2g} / E'$ $\approx$ 383~cm$^{-1}$) Raman mode. $A$, $B$, and $C$ denote the excitonic transitions. Different thickness of the SiO$_2$ (50~nm or 90~nm) are indicated. For the freestanding bilayer sample the Raman intensity of the surrounding substrate was used to calculate the intensity ratio. The lines connecting the data points are a guide to the eye. Data points in the range of the $A$ and $B$ excitons are taken from Ref. \onlinecite{Scheuschner2012a}}.
\end{figure}

\begin{figure}
\begin{minipage}{0.5\textwidth}
\includegraphics*[width=1.0\textwidth]{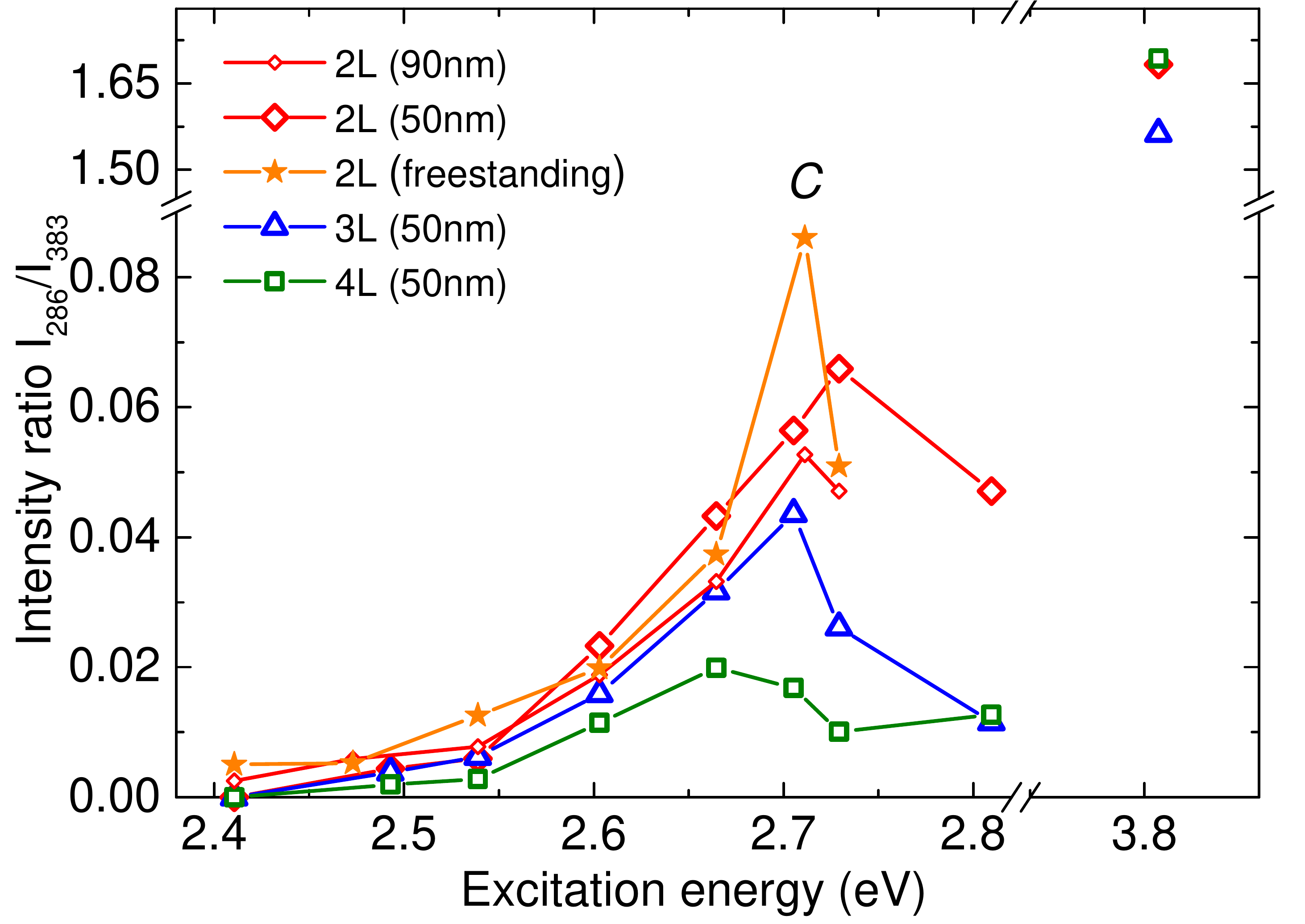}
\end{minipage}
\caption{\label{EG_ratio} Intensity ratio of the newly observed $E'$/$E_{1g}$ Raman modes of MoS$_2$ and the $E'$/$E_{1g}$ at $\approx$ 408~cm$^{-1}$ The lines connecting the data points are a guide to the eye.}
\end{figure}

As shown in Fig.\,\ref{325}, the new Raman modes in few-layer MoS$_2$ are observed at 286\,cm$^{-1}$ and at 471\,cm$^{-1}$, in excellent agreement with our group-theory analysis (Table\,\ref{tab:apli}). By polarization-dependent measurements in (\textit{\textbf{x}},\textit{\textbf{x}}) and (\textit{\textbf{x}},\textit{\textbf{y}}) scattering geometry, we confirmed our assignment of the 286\,cm$^{-1}$ peak to $E_g / E^\prime$ symmetry and of the 471\,cm$^{-1}$ peak to $A_{1g} / A'_1$ symmetry.
The spectra at 2.7\,eV and 3.8\,eV excitation energy shown in Fig.\,\ref{325}\,(a) and (b), respectively, indicate different resonance behavior of the different Raman modes. Therefore, we compare in Figs.\,\ref{A1G_ratio} and \ref{EG_ratio} the intensity behavior of the always observed Raman modes with that of the newly observed modes.

Figure \,\ref{A1G_ratio} shows the intensity of the $A_{1g} / A'_1$ mode (407\,cm$^{-1}$) in single and bilayer MoS$_2$, normalized to the number of layers and to the Raman intensity of the Si 521\,cm$^{-1}$ peak of the underlying substrate, as a function of excitation energy. As expected, the MoS$_{2}$ Raman signal is resonantly enhanced at the $A$ and $B$ excitonic transitions.\cite{Scheuschner2012a} Above 2.4\,eV excitation energy, the Raman intensity increases even further, which we attribute to resonance with the optical absorption around the $C$ exciton transition. This optical absorption peak has been recently determined to be at 2.84\,eV in single-layer MoS$_2$ and at 2.73\,eV in bilayer MoS$_2$\cite{Dhakal2014}; the intensity increase towards 2.7-2.8\,eV, as shown in Fig.\,\ref{A1G_ratio}, is in good agreement with these values. The second typically observed Raman mode, the $E_g / E^\prime$ mode at 383\,cm$^{-1}$, shows a very similar intensity dependence on the excitation energy, as shown for 1L in Fig.\,\ref{A1G_ratio}. In order to check for possible effects of the substrate in the excitation-energy dependence of the Raman intensity, \textit{e.g.}, due to interference in the SiO$_2$ layer, we measured 1L and 2L MoS$_2$ on Si substrates with SiO$_2$ thicknesses of 50\,nm and 90\,nm, as well as freestanding 2L MoS$_2$. Although the intensity ratios to the Si Raman intensity vary slightly, the overall intensity increase towards the $C$ exciton is similar in all samples. 

The newly observed modes, on the other hand, start to appear in the spectra only above 2.4\,eV excitation energy. Their \emph{relative} enhancement towards the $C$ absorption peak, however, is even stronger than that of the other Raman modes. This is shown in Fig.\,\ref{EG_ratio}, where the intensity ratio between the  ``new''  $E_g / E^\prime$ mode at 286\,cm$^{-1}$ and the peak at 383\,cm$^{-1}$ of the same symmetry is depicted for 2L, 3L, and 4L MoS$_2$. In 2L MoS$_2$, this relative intensity shows a maximum at $\approx$ 2.7\,eV; in 3L and 4L MoS$_2$ this maximum gradually shifts to lower excitation energy. This fits again nicely with the layer-number dependence of the $C$ absorption peak\cite{Dhakal2014} and supports our assignment to this resonance. At 3.8\,eV excitation energy, the ``new'' Raman modes show even higher intensity than the other, typically observed modes, see also Fig.~\ref{325}. The other newly observed mode ($A_{1g} / A_1^\prime$ at 471\,cm$^{-1}$) shows a similar trend. However, it overlaps with the second-order Raman spectrum of the LA and LA$^\prime$ phonons from the \textit{M} point\cite{Livneh2014}, which makes an analysis of the intensity less exact. In monolayer MoS$_2$ the new modes are not observed, independent of excitation energy, and in agreement with the symmetry considerations discussed above. 

In order to understand the distinct resonance behavior of the newly observed Raman modes, we consider the spatial distribution of the excitonic wave functions in few-layer MoS$_2$. Theoretical calculations and experimental results predict the wave function of the $A$ exciton in few-layer MoS$_2$ to be strongly confined to a single layer with only small overlap to the neighboring layers.\cite{Molina-Sanchez2013,Schuller2013}  Recent calculations of few-layer MoSe$_2$, which is quite similar in electronic structure to MoS$_2$, also showed that the wave functions of the conduction and valence band states at the \textit{K} point are strongly confined to a single layer.\cite{Bradley2015}

As we will discuss below, our results suggest that the $C$ exciton, on the other hand, is much less confined to a single layer in few-layer MoS$_2$, but instead extends over the entire few-layer thickness. This is in agreement with the significantly larger blueshift of the $C$ absorption peak compared to the $A$ and $B$ absorption peaks when the number of layers is reduced \cite{Dhakal2014}.

\begin{figure}
\begin{minipage}{0.5\textwidth}
\includegraphics*[width=1.0\textwidth]{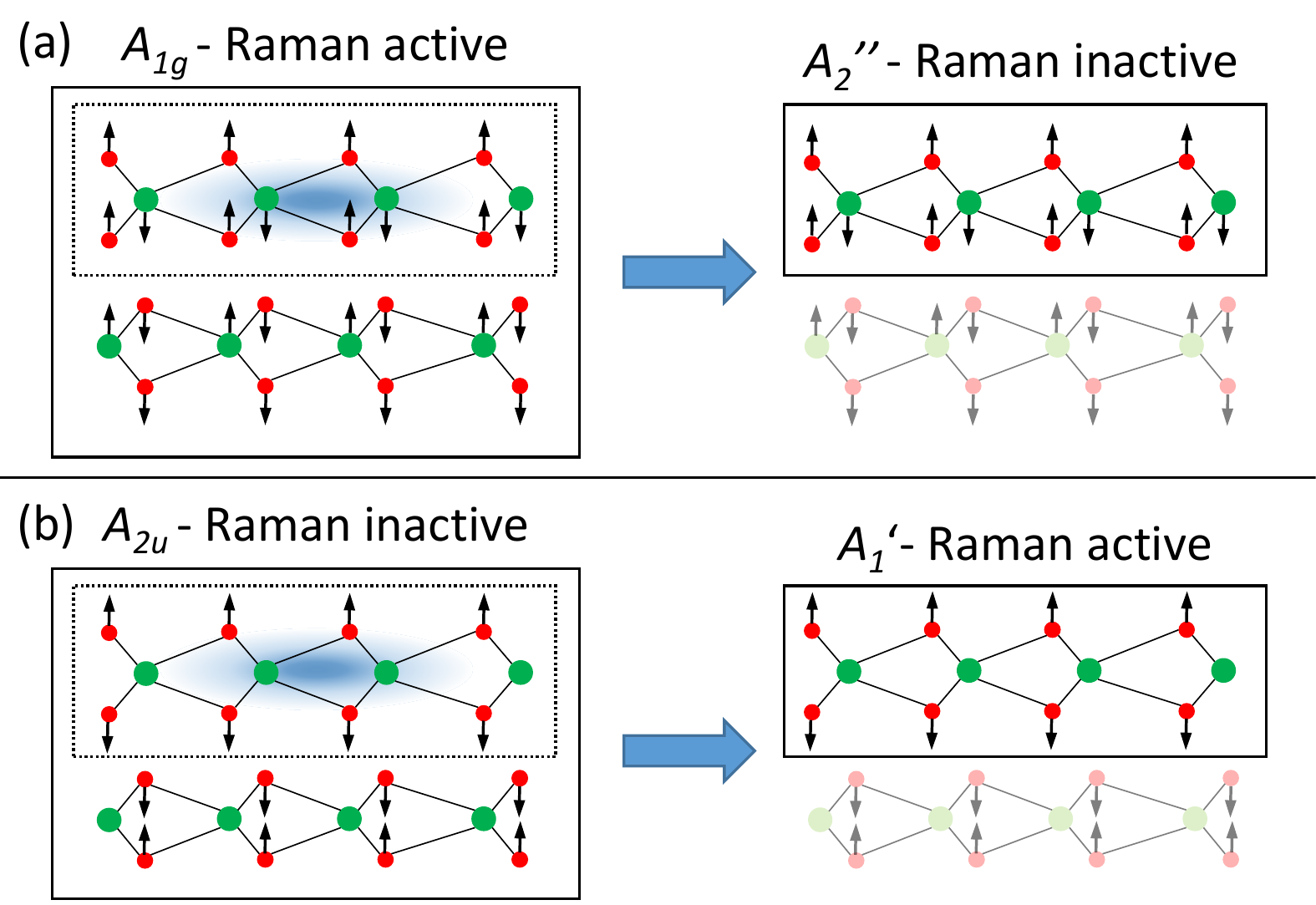}
\end{minipage}
\caption{\label{A_res} In resonance with the $A$ exciton, the Raman active $A_{1g}$ mode of 2L MoS$_2$ can be ``seen'' like the Raman inactive $A^{\prime\prime}_{2}$ mode of 1L MoS$_2$ (a). The Raman inactive $A_{2u}$ mode of 2L MoS$_2$ can be ``seen'' like a $A_{1g}$ symmetry mode of 1L MoS$_2$ (b) and starts to become observable.}
\end{figure}

The spatial confinement of the wave functions has strong implications on the Raman scattering process. If the Raman excitation is in resonance with the $A$ or $B$ exciton in few-layer MoS$_2$ the excitation is still confined mainly to a single layer [see Fig.~\ref{A_res}(a)]. As a result, the $N$-layer system reacts rather like an independent superposition of $N$ single layers from a selection-rule point of view .
In the single layer, the vibrations at 286\,cm$^{-1}$ and 471\,cm$^{-1}$ are symmetry-forbidden in backscattering geometry, and therefore their corresponding symmetry-allowed Raman signal in the few-layer system is still vanishingly small in resonance with the $A$ or $B$ exciton. Only when the optical excitation creates an exciton that is extended over the entire number of layers, as in resonance with the $C$ transition, the symmetry of the few-layer system becomes dominant and determines the selection rules, allowing the newly observed Raman modes and bringing them into resonance. 

\begin{figure}
\begin{minipage}{0.5\textwidth}
\includegraphics*[width=1.0\textwidth]{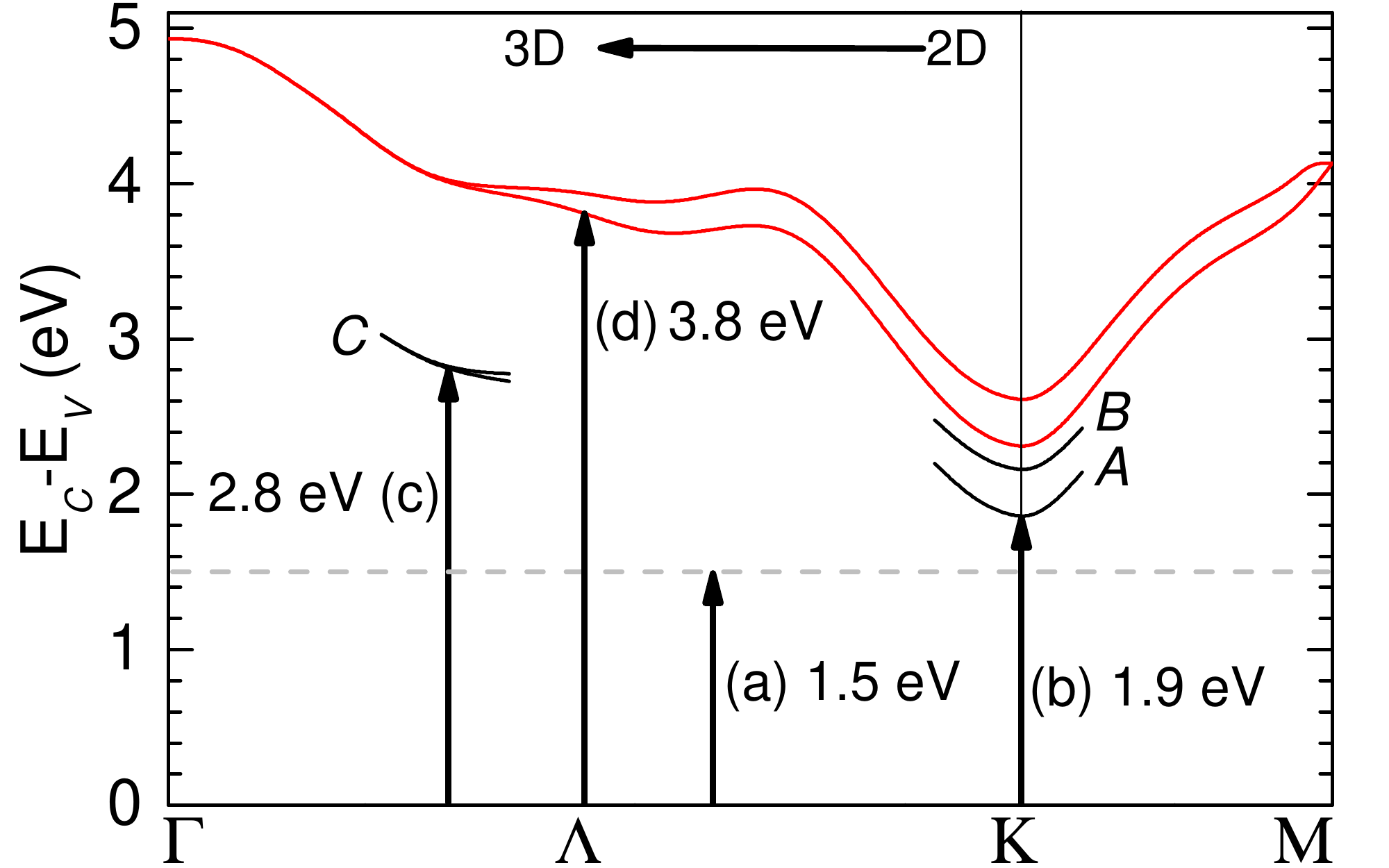}
\end{minipage}
\caption{\label{Band} Schematic views of the optical absorption process (black arrow) for (a) 1.5~eV, (b) 1.9~eV, (c) 2.8~eV and (d) 3.8~eV excitation energy. The red lines show the difference in energy of the conduction band (E$_c$) and valence band (E$_v$) states, taking the spin-orbit splitting of the conduction band into account. The black lines indicate the exciton ground states. The gray dashed line indicates a virtual transition.}
\end{figure}

The different regimes of optical excitations are schematically illustrated in Fig.\,\ref{Band}. 
We show the difference between valence-band and conduction-band energies schematically from a density-functional theory (DFT) calculation of the single-layer MoS$_2$ band structure. Because the DFT results underestimate the band gap, we applied a constant scaling factor to the energies such that the $A$ exciton is at 1.9\,eV when assuming a binding energy of 445\,meV.\cite{Berghauser2014} Figure\,\ref{Band}\,(a) illustrates a non-resonant excitation at 1.5\,eV, which leads to very weak Raman signal. We found it to be even weaker than the Si second-order modes around 300\,cm$^{-1}$ (not shown here).

At 1.9\,eV [Fig.\,\ref{Band}(b)], the $A$ exciton resonance is reached; because of the intralayer confinement, the Raman spectra of few layers and single layers are rather similar. Note, however, that the van-der-Waals and electronic interactions between the layers do exist and lead to the observed shifts in the Raman frequencies.\cite{Lee2010,Molina-Sanchez2011}
At $\approx$ 2.8\,eV [Fig.\,\ref{Band}(c)], the $C$ exciton is reached, which appears to be the first resonance which is extended over all layers. In this case, the full symmetry of the few-layer system is ``seen'', and all symmetry-allowed vibrational modes become resonant (Table \ref{tab:apli}), in particular the ones that are not Raman active in the single layer or bulk, \textit{i.e.} the new modes at 286\,cm$^{-1}$ and 471\,cm$^{-1}$.

Above this energy, we observe the same situation, \textit{i.e.}, the optical excitation is not any more confined to individual layers. Figure\,\ref{Band}(d) indicates a possible resonance for 3.8\,eV excitation energy, which might be in the continuum of optical excitation at the nearly flat bands between the $\Gamma$ and \textit{K} point of the Brillouin zone. 

Our results suggest that a high intensity of those Raman modes that are specifically only allowed in few-layer systems (compared to their single-layer or bulk counterpart) in general indicates an optical excitation that is not confined to a single layer in the few-layer system.
{\it Vice versa}, the absence of these modes indicates $(i)$ a true single-layer sample or $(ii)$ a few-layer sample where only intralayer-confined states have been excited.

Examples supporting our interpretation can be found in literature: In few-layer WSe$_2$, where the $A$ exciton energy is 1.6-1.7~eV, the new $A_{1g} / A_1^\prime$ mode has been reported for excitation energies larger 2.33\,eV. \cite{Tonndorf2013,Luo2014,Terrones2014} At lower excitation energies, the new mode was not detectable.\cite{Terrones2014} In few-layer MoTe$_2$, the new $A_{1g} / A_1^\prime$ mode has been reported for excitation energies of 1.96\,eV and 2.33\,eV, again in both cases the excitation energy are much larger than the optical band gap of <1.1~eV.\cite{Yamamoto2014,Ruppert2014}

Finally, resonant excitation into intralayer-confined excitons in few-layer systems can by itself ``activate'' Raman-inactive modes [see Fig.~\ref{A_res}(b)]: As discussed above, then the selection rules follow more the rules of $N$ independent layers. Thus, $N$ Raman peaks corresponding to a given single-layer Raman mode can be observed -- given that the frequency splitting between them can be spectrally resolved --, although only every other mode of the set of $N$ modes is Raman active in the $N$-layer system (Table\,\ref{tab:apli}). In this case, the truly Raman-allowed modes still dominate the spectra; the $N$-layer inactive but 1L-active modes are significantly weaker in intensity. Such Raman spectra have been recently observed in WS$_2$, which exhibits particularly large frequency splitting for the optical $A_{1g} / A_1^\prime$ modes.\cite{Staiger2015}

\section{Conclusion}

In conclusion, we have presented a generalized symmetry-based analysis of vibrational modes in $N$-layer materials. We have shown that the vibrational spectra in few-layer systems can have distinct signatures for the interaction between the layers compared to the corresponding single-layer or bulk material. Moreover, we have shown how to derive the vibrational modes, their symmetry, displacement vectors, and approximate frequency range from the symmetry of the single-layer system.

In general, two symmetry-related effects determine the vibrational spectra of few-layer systems: $(i)$ the symmetry group of the few-layer system can be different from the single-layer or bulk symmetry group. This can lead to activation of vibrational modes, such as the silent bulk $B_{2g}$ mode in the $D_{6h}$ symmetry group of MoS$_2$, which becomes a Raman active $A_{1g}$ mode in $N$ layers with even $N$ (symmetry group $D_{3d}$). $(ii)$ each vibration of the single layer will always lead to several combination modes of the few-layer system that show opposite behavior under inversion (horizontal reflection). As a result the combination modes resulting from Raman inactive vibrations of the single layer can be combined into Raman active vibrations of the few-layer system, such as the symmetric breathing-like vibrations arising from Raman-inactive acoustic modes, or the optical $A_2^{\prime\prime}$ mode in single-layer MoS$_2$ which becomes the Raman-active $A_1^\prime$ mode in $N$ layers with odd $N$. 
Our experimental results of such additional Raman modes in few-layer MoS$_2$ confirm our theoretical predictions. 

Based on the strong resonance of the few-layer specific Raman modes with the $C$ optical transition, we predict that the corresponding exciton wave function is extended over all layers of the few-layer MoS$_2$, in contrast to the $A$ and $B$ excitons, which appear strongly confined to a single layer. Consequently, the intensity of ``new'' few-layer Raman modes in layered materials in general can indicate the spatial distribution of the optical excitation.
The symmetry-based analysis and the resulting predictions can in general be applied to any layered crystal structure with inversion symmetry and/or horizontal reflection symmetry.

\section{Acknowledgements}

This work was supported by the European Research Council (ERC) under grant no. 259286 and by the SPP 1459 Graphene of the Deutsche Forschungsgemeinschaft (DFG).


\begin{thebibliography}{49}%
\makeatletter
\providecommand \@ifxundefined [1]{%
 \@ifx{#1\undefined}
}%
\providecommand \@ifnum [1]{%
 \ifnum #1\expandafter \@firstoftwo
 \else \expandafter \@secondoftwo
 \fi
}%
\providecommand \@ifx [1]{%
 \ifx #1\expandafter \@firstoftwo
 \else \expandafter \@secondoftwo
 \fi
}%
\providecommand \natexlab [1]{#1}%
\providecommand \enquote  [1]{``#1''}%
\providecommand \bibnamefont  [1]{#1}%
\providecommand \bibfnamefont [1]{#1}%
\providecommand \citenamefont [1]{#1}%
\providecommand \href@noop [0]{\@secondoftwo}%
\providecommand \href [0]{\begingroup \@sanitize@url \@href}%
\providecommand \@href[1]{\@@startlink{#1}\@@href}%
\providecommand \@@href[1]{\endgroup#1\@@endlink}%
\providecommand \@sanitize@url [0]{\catcode `\\12\catcode `\$12\catcode
  `\&12\catcode `\#12\catcode `\^12\catcode `\_12\catcode `\%12\relax}%
\providecommand \@@startlink[1]{}%
\providecommand \@@endlink[0]{}%
\providecommand \url  [0]{\begingroup\@sanitize@url \@url }%
\providecommand \@url [1]{\endgroup\@href {#1}{\urlprefix }}%
\providecommand \urlprefix  [0]{URL }%
\providecommand \Eprint [0]{\href }%
\providecommand \doibase [0]{http://dx.doi.org/}%
\providecommand \selectlanguage [0]{\@gobble}%
\providecommand \bibinfo  [0]{\@secondoftwo}%
\providecommand \bibfield  [0]{\@secondoftwo}%
\providecommand \translation [1]{[#1]}%
\providecommand \BibitemOpen [0]{}%
\providecommand \bibitemStop [0]{}%
\providecommand \bibitemNoStop [0]{.\EOS\space}%
\providecommand \EOS [0]{\spacefactor3000\relax}%
\providecommand \BibitemShut  [1]{\csname bibitem#1\endcsname}%
\let\auto@bib@innerbib\@empty
\bibitem [{\citenamefont {Nicolosi}\ \emph {et~al.}(2013)\citenamefont
  {Nicolosi}, \citenamefont {Chhowalla}, \citenamefont {Kanatzidis},
  \citenamefont {Strano},\ and\ \citenamefont {Coleman}}]{Nicolosi2013}%
  \BibitemOpen
  \bibfield  {author} {\bibinfo {author} {\bibfnamefont {V.}~\bibnamefont
  {Nicolosi}}, \bibinfo {author} {\bibfnamefont {M.}~\bibnamefont {Chhowalla}},
  \bibinfo {author} {\bibfnamefont {M.~G.}\ \bibnamefont {Kanatzidis}},
  \bibinfo {author} {\bibfnamefont {M.~S.}\ \bibnamefont {Strano}}, \ and\
  \bibinfo {author} {\bibfnamefont {J.~N.}\ \bibnamefont {Coleman}},\ }\href
  {\doibase 10.1126/science.1226419} {\bibfield  {journal} {\bibinfo  {journal}
  {Science}\ }\textbf {\bibinfo {volume} {340}},\ \bibinfo {pages} {1226419}
  (\bibinfo {year} {2013})}\BibitemShut {NoStop}%
\bibitem [{\citenamefont {Geim}\ and\ \citenamefont
  {Grigorieva}(2013)}]{Geim2013}%
  \BibitemOpen
  \bibfield  {author} {\bibinfo {author} {\bibfnamefont {A.~K.}\ \bibnamefont
  {Geim}}\ and\ \bibinfo {author} {\bibfnamefont {I.~V.}\ \bibnamefont
  {Grigorieva}},\ }\href {\doibase 10.1038/nature12385} {\bibfield  {journal}
  {\bibinfo  {journal} {Nature}\ }\textbf {\bibinfo {volume} {499}},\ \bibinfo
  {pages} {419} (\bibinfo {year} {2013})}\BibitemShut {NoStop}%
\bibitem [{\citenamefont {Radisavljevic}\ \emph {et~al.}(2011)\citenamefont
  {Radisavljevic}, \citenamefont {Radenovic}, \citenamefont {Brivio},
  \citenamefont {Giacometti},\ and\ \citenamefont {Kis}}]{Radisavljevic2011}%
  \BibitemOpen
  \bibfield  {author} {\bibinfo {author} {\bibfnamefont {B.}~\bibnamefont
  {Radisavljevic}}, \bibinfo {author} {\bibfnamefont {A.}~\bibnamefont
  {Radenovic}}, \bibinfo {author} {\bibfnamefont {J.}~\bibnamefont {Brivio}},
  \bibinfo {author} {\bibfnamefont {V.}~\bibnamefont {Giacometti}}, \ and\
  \bibinfo {author} {\bibfnamefont {A.}~\bibnamefont {Kis}},\ }\href {\doibase
  10.1038/nnano.2010.279} {\bibfield  {journal} {\bibinfo  {journal} {Nature
  Nanotechnology}\ }\textbf {\bibinfo {volume} {6}},\ \bibinfo {pages} {147}
  (\bibinfo {year} {2011})}\BibitemShut {NoStop}%
\bibitem [{\citenamefont {Lembke}\ and\ \citenamefont
  {Kis}(2012)}]{Lembke2012}%
  \BibitemOpen
  \bibfield  {author} {\bibinfo {author} {\bibfnamefont {D.}~\bibnamefont
  {Lembke}}\ and\ \bibinfo {author} {\bibfnamefont {A.}~\bibnamefont {Kis}},\
  }\href {\doibase 10.1021/nn303772b} {\bibfield  {journal} {\bibinfo
  {journal} {ACS Nano}\ }\textbf {\bibinfo {volume} {6}},\ \bibinfo {pages}
  {10070} (\bibinfo {year} {2012})}\BibitemShut {NoStop}%
\bibitem [{\citenamefont {Buscema}\ \emph {et~al.}(2013)\citenamefont
  {Buscema}, \citenamefont {Barkelid}, \citenamefont {Zwiller}, \citenamefont
  {van~der Zant}, \citenamefont {Steele},\ and\ \citenamefont
  {Castellanos-Gomez}}]{Buscema2013}%
  \BibitemOpen
  \bibfield  {author} {\bibinfo {author} {\bibfnamefont {M.}~\bibnamefont
  {Buscema}}, \bibinfo {author} {\bibfnamefont {M.}~\bibnamefont {Barkelid}},
  \bibinfo {author} {\bibfnamefont {V.}~\bibnamefont {Zwiller}}, \bibinfo
  {author} {\bibfnamefont {H.~S.~J.}\ \bibnamefont {van~der Zant}}, \bibinfo
  {author} {\bibfnamefont {G.~A.}\ \bibnamefont {Steele}}, \ and\ \bibinfo
  {author} {\bibfnamefont {A.}~\bibnamefont {Castellanos-Gomez}},\ }\href
  {\doibase 10.1021/nl303321g} {\bibfield  {journal} {\bibinfo  {journal} {Nano
  Letters}\ }\textbf {\bibinfo {volume} {13}},\ \bibinfo {pages} {358}
  (\bibinfo {year} {2013})}\BibitemShut {NoStop}%
\bibitem [{\citenamefont {Bertolazzi}\ \emph {et~al.}(2011)\citenamefont
  {Bertolazzi}, \citenamefont {Brivio},\ and\ \citenamefont
  {Kis}}]{Bertolazzi2011a}%
  \BibitemOpen
  \bibfield  {author} {\bibinfo {author} {\bibfnamefont {S.}~\bibnamefont
  {Bertolazzi}}, \bibinfo {author} {\bibfnamefont {J.}~\bibnamefont {Brivio}},
  \ and\ \bibinfo {author} {\bibfnamefont {A.}~\bibnamefont {Kis}},\ }\href
  {\doibase 10.1021/nn203879f} {\bibfield  {journal} {\bibinfo  {journal} {ACS
  Nano}\ }\textbf {\bibinfo {volume} {5}},\ \bibinfo {pages} {9703} (\bibinfo
  {year} {2011})}\BibitemShut {NoStop}%
\bibitem [{\citenamefont {Ochedowski}\ \emph {et~al.}(2014)\citenamefont
  {Ochedowski}, \citenamefont {Marinov}, \citenamefont {Scheuschner},
  \citenamefont {Poloczek}, \citenamefont {Bussmann}, \citenamefont
  {Maultzsch},\ and\ \citenamefont {Schleberger}}]{Ochedowski2014}%
  \BibitemOpen
  \bibfield  {author} {\bibinfo {author} {\bibfnamefont {O.}~\bibnamefont
  {Ochedowski}}, \bibinfo {author} {\bibfnamefont {K.}~\bibnamefont {Marinov}},
  \bibinfo {author} {\bibfnamefont {N.}~\bibnamefont {Scheuschner}}, \bibinfo
  {author} {\bibfnamefont {A.}~\bibnamefont {Poloczek}}, \bibinfo {author}
  {\bibfnamefont {B.~K.}\ \bibnamefont {Bussmann}}, \bibinfo {author}
  {\bibfnamefont {J.}~\bibnamefont {Maultzsch}}, \ and\ \bibinfo {author}
  {\bibfnamefont {M.}~\bibnamefont {Schleberger}},\ }\href {\doibase
  10.3762/bjnano.5.32} {\bibfield  {journal} {\bibinfo  {journal} {Beilstein
  Journal of Nanotechnology}\ }\textbf {\bibinfo {volume} {5}},\ \bibinfo
  {pages} {291} (\bibinfo {year} {2014})}\BibitemShut {NoStop}%
\bibitem [{\citenamefont {Mak}\ \emph {et~al.}(2010)\citenamefont {Mak},
  \citenamefont {Lee}, \citenamefont {Hone}, \citenamefont {Shan},\ and\
  \citenamefont {Heinz}}]{Mak2010}%
  \BibitemOpen
  \bibfield  {author} {\bibinfo {author} {\bibfnamefont {K.~F.}\ \bibnamefont
  {Mak}}, \bibinfo {author} {\bibfnamefont {C.}~\bibnamefont {Lee}}, \bibinfo
  {author} {\bibfnamefont {J.}~\bibnamefont {Hone}}, \bibinfo {author}
  {\bibfnamefont {J.}~\bibnamefont {Shan}}, \ and\ \bibinfo {author}
  {\bibfnamefont {T.~F.}\ \bibnamefont {Heinz}},\ }\href {\doibase
  10.1103/PhysRevLett.105.136805} {\bibfield  {journal} {\bibinfo  {journal}
  {Physical Review Letters}\ }\textbf {\bibinfo {volume} {105}},\ \bibinfo
  {pages} {136805} (\bibinfo {year} {2010})}\BibitemShut {NoStop}%
\bibitem [{\citenamefont {Tonndorf}\ \emph {et~al.}(2013)\citenamefont
  {Tonndorf}, \citenamefont {Schmidt}, \citenamefont {B\"{o}ttger},
  \citenamefont {Zhang}, \citenamefont {B\"{o}rner}, \citenamefont {Liebig},
  \citenamefont {Albrecht}, \citenamefont {Kloc}, \citenamefont {Gordan},
  \citenamefont {Zahn}, \citenamefont {{Michaelis de Vasconcellos}},\ and\
  \citenamefont {Bratschitsch}}]{Tonndorf2013}%
  \BibitemOpen
  \bibfield  {author} {\bibinfo {author} {\bibfnamefont {P.}~\bibnamefont
  {Tonndorf}}, \bibinfo {author} {\bibfnamefont {R.}~\bibnamefont {Schmidt}},
  \bibinfo {author} {\bibfnamefont {P.}~\bibnamefont {B\"{o}ttger}}, \bibinfo
  {author} {\bibfnamefont {X.}~\bibnamefont {Zhang}}, \bibinfo {author}
  {\bibfnamefont {J.}~\bibnamefont {B\"{o}rner}}, \bibinfo {author}
  {\bibfnamefont {A.}~\bibnamefont {Liebig}}, \bibinfo {author} {\bibfnamefont
  {M.}~\bibnamefont {Albrecht}}, \bibinfo {author} {\bibfnamefont
  {C.}~\bibnamefont {Kloc}}, \bibinfo {author} {\bibfnamefont {O.}~\bibnamefont
  {Gordan}}, \bibinfo {author} {\bibfnamefont {D.~R.~T.}\ \bibnamefont {Zahn}},
  \bibinfo {author} {\bibfnamefont {S.}~\bibnamefont {{Michaelis de
  Vasconcellos}}}, \ and\ \bibinfo {author} {\bibfnamefont {R.}~\bibnamefont
  {Bratschitsch}},\ }\href {\doibase http://dx.doi.org/10.1364/OE.21.004908}
  {\bibfield  {journal} {\bibinfo  {journal} {Optics Express}\ }\textbf
  {\bibinfo {volume} {21}},\ \bibinfo {pages} {4908} (\bibinfo {year}
  {2013})}\BibitemShut {NoStop}%
\bibitem [{\citenamefont {Livneh}\ and\ \citenamefont
  {Sterer}(2010)}]{Livneh2010}%
  \BibitemOpen
  \bibfield  {author} {\bibinfo {author} {\bibfnamefont {T.}~\bibnamefont
  {Livneh}}\ and\ \bibinfo {author} {\bibfnamefont {E.}~\bibnamefont
  {Sterer}},\ }\href {\doibase 10.1103/PhysRevB.81.195209} {\bibfield
  {journal} {\bibinfo  {journal} {Physical Review B}\ }\textbf {\bibinfo
  {volume} {81}},\ \bibinfo {pages} {195209} (\bibinfo {year}
  {2010})}\BibitemShut {NoStop}%
\bibitem [{\citenamefont {Scheuschner}\ \emph {et~al.}(2014)\citenamefont
  {Scheuschner}, \citenamefont {Ochedowski}, \citenamefont {Kaulitz},
  \citenamefont {Gillen}, \citenamefont {Schleberger},\ and\ \citenamefont
  {Maultzsch}}]{Scheuschner2014}%
  \BibitemOpen
  \bibfield  {author} {\bibinfo {author} {\bibfnamefont {N.}~\bibnamefont
  {Scheuschner}}, \bibinfo {author} {\bibfnamefont {O.}~\bibnamefont
  {Ochedowski}}, \bibinfo {author} {\bibfnamefont {A.-M.}\ \bibnamefont
  {Kaulitz}}, \bibinfo {author} {\bibfnamefont {R.}~\bibnamefont {Gillen}},
  \bibinfo {author} {\bibfnamefont {M.}~\bibnamefont {Schleberger}}, \ and\
  \bibinfo {author} {\bibfnamefont {J.}~\bibnamefont {Maultzsch}},\ }\href
  {\doibase 10.1103/PhysRevB.89.125406} {\bibfield  {journal} {\bibinfo
  {journal} {Physical Review B}\ }\textbf {\bibinfo {volume} {89}},\ \bibinfo
  {pages} {125406} (\bibinfo {year} {2014})}\BibitemShut {NoStop}%
\bibitem [{\citenamefont {Conley}\ \emph {et~al.}(2013)\citenamefont {Conley},
  \citenamefont {Wang}, \citenamefont {Ziegler}, \citenamefont {Haglund},
  \citenamefont {Pantelides},\ and\ \citenamefont {Bolotin}}]{Conley2013a}%
  \BibitemOpen
  \bibfield  {author} {\bibinfo {author} {\bibfnamefont {H.~J.}\ \bibnamefont
  {Conley}}, \bibinfo {author} {\bibfnamefont {B.}~\bibnamefont {Wang}},
  \bibinfo {author} {\bibfnamefont {J.~I.}\ \bibnamefont {Ziegler}}, \bibinfo
  {author} {\bibfnamefont {R.~F.}\ \bibnamefont {Haglund}}, \bibinfo {author}
  {\bibfnamefont {S.~T.}\ \bibnamefont {Pantelides}}, \ and\ \bibinfo {author}
  {\bibfnamefont {K.~I.}\ \bibnamefont {Bolotin}},\ }\href {\doibase
  10.1021/nl4014748} {\bibfield  {journal} {\bibinfo  {journal} {Nano Letters}\
  }\textbf {\bibinfo {volume} {13}},\ \bibinfo {pages} {3626} (\bibinfo {year}
  {2013})}\BibitemShut {NoStop}%
\bibitem [{\citenamefont {He}\ \emph {et~al.}(2013)\citenamefont {He},
  \citenamefont {Poole}, \citenamefont {Mak},\ and\ \citenamefont
  {Shan}}]{He2013}%
  \BibitemOpen
  \bibfield  {author} {\bibinfo {author} {\bibfnamefont {K.}~\bibnamefont
  {He}}, \bibinfo {author} {\bibfnamefont {C.}~\bibnamefont {Poole}}, \bibinfo
  {author} {\bibfnamefont {K.~F.}\ \bibnamefont {Mak}}, \ and\ \bibinfo
  {author} {\bibfnamefont {J.}~\bibnamefont {Shan}},\ }\href {\doibase
  10.1021/nl4013166} {\bibfield  {journal} {\bibinfo  {journal} {Nano Letters}\
  }\textbf {\bibinfo {volume} {13}},\ \bibinfo {pages} {2931} (\bibinfo {year}
  {2013})}\BibitemShut {NoStop}%
\bibitem [{\citenamefont {Chakraborty}\ \emph {et~al.}(2012)\citenamefont
  {Chakraborty}, \citenamefont {Bera}, \citenamefont {Muthu}, \citenamefont
  {Bhowmick}, \citenamefont {Waghmare},\ and\ \citenamefont
  {Sood}}]{Chakraborty2012}%
  \BibitemOpen
  \bibfield  {author} {\bibinfo {author} {\bibfnamefont {B.}~\bibnamefont
  {Chakraborty}}, \bibinfo {author} {\bibfnamefont {A.}~\bibnamefont {Bera}},
  \bibinfo {author} {\bibfnamefont {D.~V.~S.}\ \bibnamefont {Muthu}}, \bibinfo
  {author} {\bibfnamefont {S.}~\bibnamefont {Bhowmick}}, \bibinfo {author}
  {\bibfnamefont {U.~V.}\ \bibnamefont {Waghmare}}, \ and\ \bibinfo {author}
  {\bibfnamefont {A.~K.}\ \bibnamefont {Sood}},\ }\href {\doibase
  10.1103/PhysRevB.85.161403} {\bibfield  {journal} {\bibinfo  {journal}
  {Physical Review B}\ }\textbf {\bibinfo {volume} {85}},\ \bibinfo {pages}
  {161403} (\bibinfo {year} {2012})}\BibitemShut {NoStop}%
\bibitem [{\citenamefont {Molina-S\'{a}nchez}\ and\ \citenamefont
  {Wirtz}(2011)}]{Molina-Sanchez2011}%
  \BibitemOpen
  \bibfield  {author} {\bibinfo {author} {\bibfnamefont {A.}~\bibnamefont
  {Molina-S\'{a}nchez}}\ and\ \bibinfo {author} {\bibfnamefont
  {L.}~\bibnamefont {Wirtz}},\ }\href {\doibase 10.1103/PhysRevB.84.155413}
  {\bibfield  {journal} {\bibinfo  {journal} {Physical Review B}\ }\textbf
  {\bibinfo {volume} {84}},\ \bibinfo {pages} {155413} (\bibinfo {year}
  {2011})}\BibitemShut {NoStop}%
\bibitem [{\citenamefont {Lee}\ \emph {et~al.}(2010{\natexlab{a}})\citenamefont
  {Lee}, \citenamefont {Yan}, \citenamefont {Brus}, \citenamefont {Heinz},
  \citenamefont {Hone},\ and\ \citenamefont {Ryu}}]{Lee2010c}%
  \BibitemOpen
  \bibfield  {author} {\bibinfo {author} {\bibfnamefont {C.}~\bibnamefont
  {Lee}}, \bibinfo {author} {\bibfnamefont {H.}~\bibnamefont {Yan}}, \bibinfo
  {author} {\bibfnamefont {L.}~\bibnamefont {Brus}}, \bibinfo {author}
  {\bibfnamefont {T.}~\bibnamefont {Heinz}}, \bibinfo {author} {\bibfnamefont
  {J.}~\bibnamefont {Hone}}, \ and\ \bibinfo {author} {\bibfnamefont
  {S.}~\bibnamefont {Ryu}},\ }\href
  {http://pubs.acs.org/doi/abs/10.1021/nn1003937} {\bibfield  {journal}
  {\bibinfo  {journal} {ACS Nano}\ }\textbf {\bibinfo {volume} {4}},\ \bibinfo
  {pages} {2695} (\bibinfo {year} {2010}{\natexlab{a}})}\BibitemShut {NoStop}%
\bibitem [{\citenamefont {Yamamoto}\ \emph {et~al.}(2014)\citenamefont
  {Yamamoto}, \citenamefont {Wang}, \citenamefont {Ni}, \citenamefont {Lin},
  \citenamefont {Li}, \citenamefont {Aikawa}, \citenamefont {Jian},
  \citenamefont {Ueno}, \citenamefont {Wakabayashi},\ and\ \citenamefont
  {Tsukagoshi}}]{Yamamoto2014}%
  \BibitemOpen
  \bibfield  {author} {\bibinfo {author} {\bibfnamefont {M.}~\bibnamefont
  {Yamamoto}}, \bibinfo {author} {\bibfnamefont {S.~T.}\ \bibnamefont {Wang}},
  \bibinfo {author} {\bibfnamefont {M.}~\bibnamefont {Ni}}, \bibinfo {author}
  {\bibfnamefont {Y.-F.}\ \bibnamefont {Lin}}, \bibinfo {author} {\bibfnamefont
  {S.-L.}\ \bibnamefont {Li}}, \bibinfo {author} {\bibfnamefont
  {S.}~\bibnamefont {Aikawa}}, \bibinfo {author} {\bibfnamefont {W.-B.}\
  \bibnamefont {Jian}}, \bibinfo {author} {\bibfnamefont {K.}~\bibnamefont
  {Ueno}}, \bibinfo {author} {\bibfnamefont {K.}~\bibnamefont {Wakabayashi}}, \
  and\ \bibinfo {author} {\bibfnamefont {K.}~\bibnamefont {Tsukagoshi}},\
  }\href {\doibase 10.1021/nn5007607} {\bibfield  {journal} {\bibinfo
  {journal} {ACS Nano}\ }\textbf {\bibinfo {volume} {8}},\ \bibinfo {pages}
  {3895} (\bibinfo {year} {2014})}\BibitemShut {NoStop}%
\bibitem [{\citenamefont {Terrones}\ \emph {et~al.}(2014)\citenamefont
  {Terrones}, \citenamefont {Corro}, \citenamefont {Feng}, \citenamefont
  {Poumirol}, \citenamefont {Rhodes}, \citenamefont {Smirnov}, \citenamefont
  {Pradhan}, \citenamefont {Lin}, \citenamefont {Nguyen}, \citenamefont
  {El\'{\i}as}, \citenamefont {Mallouk}, \citenamefont {Balicas}, \citenamefont
  {Pimenta},\ and\ \citenamefont {Terrones}}]{Terrones2014}%
  \BibitemOpen
  \bibfield  {author} {\bibinfo {author} {\bibfnamefont {H.}~\bibnamefont
  {Terrones}}, \bibinfo {author} {\bibfnamefont {E.~D.}\ \bibnamefont {Corro}},
  \bibinfo {author} {\bibfnamefont {S.}~\bibnamefont {Feng}}, \bibinfo {author}
  {\bibfnamefont {J.~M.}\ \bibnamefont {Poumirol}}, \bibinfo {author}
  {\bibfnamefont {D.}~\bibnamefont {Rhodes}}, \bibinfo {author} {\bibfnamefont
  {D.}~\bibnamefont {Smirnov}}, \bibinfo {author} {\bibfnamefont {N.~R.}\
  \bibnamefont {Pradhan}}, \bibinfo {author} {\bibfnamefont {Z.}~\bibnamefont
  {Lin}}, \bibinfo {author} {\bibfnamefont {M.~A.~T.}\ \bibnamefont {Nguyen}},
  \bibinfo {author} {\bibfnamefont {A.~L.}\ \bibnamefont {El\'{\i}as}},
  \bibinfo {author} {\bibfnamefont {T.~E.}\ \bibnamefont {Mallouk}}, \bibinfo
  {author} {\bibfnamefont {L.}~\bibnamefont {Balicas}}, \bibinfo {author}
  {\bibfnamefont {M.~A.}\ \bibnamefont {Pimenta}}, \ and\ \bibinfo {author}
  {\bibfnamefont {M.}~\bibnamefont {Terrones}},\ }\href {\doibase
  10.1038/srep04215} {\bibfield  {journal} {\bibinfo  {journal} {Scientific
  Reports}\ }\textbf {\bibinfo {volume} {4}},\ \bibinfo {pages} {4215}
  (\bibinfo {year} {2014})}\BibitemShut {NoStop}%
\bibitem [{\citenamefont {Luo}\ \emph {et~al.}(2013)\citenamefont {Luo},
  \citenamefont {Zhao}, \citenamefont {Zhang}, \citenamefont {Toh},
  \citenamefont {Kloc}, \citenamefont {Xiong},\ and\ \citenamefont
  {Quek}}]{Luo2014}%
  \BibitemOpen
  \bibfield  {author} {\bibinfo {author} {\bibfnamefont {X.}~\bibnamefont
  {Luo}}, \bibinfo {author} {\bibfnamefont {Y.}~\bibnamefont {Zhao}}, \bibinfo
  {author} {\bibfnamefont {J.}~\bibnamefont {Zhang}}, \bibinfo {author}
  {\bibfnamefont {M.}~\bibnamefont {Toh}}, \bibinfo {author} {\bibfnamefont
  {C.}~\bibnamefont {Kloc}}, \bibinfo {author} {\bibfnamefont {Q.}~\bibnamefont
  {Xiong}}, \ and\ \bibinfo {author} {\bibfnamefont {S.~Y.}\ \bibnamefont
  {Quek}},\ }\href {\doibase 10.1103/PhysRevB.88.195313} {\bibfield  {journal}
  {\bibinfo  {journal} {Physical Review B}\ }\textbf {\bibinfo {volume} {88}},\
  \bibinfo {pages} {195313} (\bibinfo {year} {2013})}\BibitemShut {NoStop}%
\bibitem [{\citenamefont {Hajiyev}\ \emph {et~al.}(2013)\citenamefont
  {Hajiyev}, \citenamefont {Cong}, \citenamefont {Qiu},\ and\ \citenamefont
  {Yu}}]{Hajiyev2013}%
  \BibitemOpen
  \bibfield  {author} {\bibinfo {author} {\bibfnamefont {P.}~\bibnamefont
  {Hajiyev}}, \bibinfo {author} {\bibfnamefont {C.}~\bibnamefont {Cong}},
  \bibinfo {author} {\bibfnamefont {C.}~\bibnamefont {Qiu}}, \ and\ \bibinfo
  {author} {\bibfnamefont {T.}~\bibnamefont {Yu}},\ }\href {\doibase
  10.1038/srep02593} {\bibfield  {journal} {\bibinfo  {journal} {Scientific
  Reports}\ }\textbf {\bibinfo {volume} {3}},\ \bibinfo {pages} {2593}
  (\bibinfo {year} {2013})}\BibitemShut {NoStop}%
\bibitem [{\citenamefont {Verble}\ and\ \citenamefont
  {Wieting}(1970)}]{Verble1970}%
  \BibitemOpen
  \bibfield  {author} {\bibinfo {author} {\bibfnamefont {J.}~\bibnamefont
  {Verble}}\ and\ \bibinfo {author} {\bibfnamefont {T.}~\bibnamefont
  {Wieting}},\ }\href {http://link.aps.org/doi/10.1103/PhysRevLett.25.362}
  {\bibfield  {journal} {\bibinfo  {journal} {Physical Review Letters}\
  }\textbf {\bibinfo {volume} {25}},\ \bibinfo {pages} {362} (\bibinfo {year}
  {1970})}\BibitemShut {NoStop}%
\bibitem [{\citenamefont {Ribeiro-Soares}\ \emph {et~al.}(2014)\citenamefont
  {Ribeiro-Soares}, \citenamefont {Almeida}, \citenamefont {Barros},
  \citenamefont {Araujo}, \citenamefont {Dresselhaus}, \citenamefont
  {Can\c{c}ado},\ and\ \citenamefont {Jorio}}]{Ribeiro-Soares2014}%
  \BibitemOpen
  \bibfield  {author} {\bibinfo {author} {\bibfnamefont {J.}~\bibnamefont
  {Ribeiro-Soares}}, \bibinfo {author} {\bibfnamefont {R.~M.}\ \bibnamefont
  {Almeida}}, \bibinfo {author} {\bibfnamefont {E.~B.}\ \bibnamefont {Barros}},
  \bibinfo {author} {\bibfnamefont {P.~T.}\ \bibnamefont {Araujo}}, \bibinfo
  {author} {\bibfnamefont {M.~S.}\ \bibnamefont {Dresselhaus}}, \bibinfo
  {author} {\bibfnamefont {L.~G.}\ \bibnamefont {Can\c{c}ado}}, \ and\ \bibinfo
  {author} {\bibfnamefont {A.}~\bibnamefont {Jorio}},\ }\href {\doibase
  10.1103/PhysRevB.90.115438} {\bibfield  {journal} {\bibinfo  {journal}
  {Physical Review B}\ }\textbf {\bibinfo {volume} {90}},\ \bibinfo {pages}
  {115438} (\bibinfo {year} {2014})}\BibitemShut {NoStop}%
\bibitem [{\citenamefont {Cardona}(1995)}]{CardonaII}%
  \BibitemOpen
  \bibfield  {author} {\bibinfo {author} {\bibnamefont {M. Cardona}},\ }\href@noop
  {} {\emph {\bibinfo {title} {Light scattering in Solids}}},\ \bibinfo
  {edition} {2nd}\ ed.,\ Vol.~\bibinfo {volume} {2}\ (\bibinfo  {publisher}
  {Springer-Verlag},\ \bibinfo {address} {Berlin},\ \bibinfo {year}
  {1995})\BibitemShut {NoStop}%
\bibitem [{\citenamefont {Wieting}\ and\ \citenamefont
  {Verble}(1971)}]{Wieting1971}%
  \BibitemOpen
  \bibfield  {author} {\bibinfo {author} {\bibfnamefont {T.}~\bibnamefont
  {Wieting}}\ and\ \bibinfo {author} {\bibfnamefont {J.}~\bibnamefont
  {Verble}},\ }\href {http://prb.aps.org/abstract/PRB/v3/i12/p4286\_1}
  {\bibfield  {journal} {\bibinfo  {journal} {Physical Review B}\ }\textbf {\bibinfo {volume} {3}},\ \bibinfo {pages}
  {4286}  (\bibinfo
  {year} {1971})}\BibitemShut {NoStop}%
\bibitem [{\citenamefont {Chen}\ and\ \citenamefont {Wang}(1974)}]{Chen1974}%
  \BibitemOpen
  \bibfield  {author} {\bibinfo {author} {\bibfnamefont {J.~M.}\ \bibnamefont
  {Chen}}\ and\ \bibinfo {author} {\bibfnamefont {C.~S.}\ \bibnamefont
  {Wang}},\ }\href@noop {} {\bibfield  {journal} {\bibinfo  {journal}
  {Spectrum}\ }\textbf {\bibinfo {volume} {14}},\ \bibinfo {pages} {857}
  (\bibinfo {year} {1974})}\BibitemShut {NoStop}%
\bibitem [{\citenamefont {Zhao}\ \emph {et~al.}(2013)\citenamefont {Zhao},
  \citenamefont {Luo}, \citenamefont {Li}, \citenamefont {Zhang}, \citenamefont
  {Araujo}, \citenamefont {Gan}, \citenamefont {Wu}, \citenamefont {Zhang},
  \citenamefont {Quek}, \citenamefont {Dresselhaus},\ and\ \citenamefont
  {Xiong}}]{Zhao2013}%
  \BibitemOpen
  \bibfield  {author} {\bibinfo {author} {\bibfnamefont {Y.}~\bibnamefont
  {Zhao}}, \bibinfo {author} {\bibfnamefont {X.}~\bibnamefont {Luo}}, \bibinfo
  {author} {\bibfnamefont {H.}~\bibnamefont {Li}}, \bibinfo {author}
  {\bibfnamefont {J.}~\bibnamefont {Zhang}}, \bibinfo {author} {\bibfnamefont
  {P.~T.}\ \bibnamefont {Araujo}}, \bibinfo {author} {\bibfnamefont {C.~K.}\
  \bibnamefont {Gan}}, \bibinfo {author} {\bibfnamefont {J.}~\bibnamefont
  {Wu}}, \bibinfo {author} {\bibfnamefont {H.}~\bibnamefont {Zhang}}, \bibinfo
  {author} {\bibfnamefont {S.~Y.}\ \bibnamefont {Quek}}, \bibinfo {author}
  {\bibfnamefont {M.~S.}\ \bibnamefont {Dresselhaus}}, \ and\ \bibinfo {author}
  {\bibfnamefont {Q.}~\bibnamefont {Xiong}},\ }\href {\doibase
  10.1021/nl304169w} {\bibfield  {journal} {\bibinfo  {journal} {Nano Letters}\
  }\textbf {\bibinfo {volume} {13}},\ \bibinfo {pages} {1007} (\bibinfo {year}
  {2013})}\BibitemShut {NoStop}%
\bibitem [{Note1()}]{Note1}%
  \BibitemOpen
  \bibinfo {note} {We define a set \textbf{G} of $c=\left\lceil \frac{N}{2} \right\rceil$ $N$-dimensional vectors $\textbf{\textit{g}}^{(i)}$ $\in \Re^N$ with $i=\left\{x\in \mathbb{N} | 0<x\leq c \right\}$ and the $j$-th component $\textit{g}^{(i)}_{j}$ of \textbf{\textit{g}}$^{(i)}$ as $\textit{g}^{(i)}_{j}=\delta_{i,j}+\delta_{i,N+1-j}$. All elements of \textbf{G} are orthogonal and eigenvectors to $\Sigma$ with the eigenvalue $\kappa=+1$. The component vectors of the set \textbf{G} will also act as a basis to a $c$-dimensional subspace $V_g$ of the $\Re^N$. We define a second set \textbf{U} of $f=\left\lfloor \frac{N}{2} \right\rfloor$ $N$-dimensional vectors $\textbf{\textit{u}}^{(k)}$ $\in \Re^N$ with $k=\left\{x\in \mathbb{N} | 0<x\leq f \right\}$ and the $j$-th components $\textit{u}^{(k)}_{j}$ of $\textbf{\textit{u}}^{(k)}$ as $\textit{u}^{(k)}_{j}=\delta_{k,j}-\delta_{k,N+1-j}$. Like the elements of \textbf{G}, all elements of \textbf{U} are orthogonal and eigenvectors of $\Sigma$, but with the eigenvalue $\kappa=-1$. Furthermore all elements of \textbf{U} are orthogonal to all elements of \textbf{G}. As before for \textbf{G}, the elements of \textbf{U} also act as a basis to an $f$-dimensional subspace $V_u$ of the $\Re^N$. As all elements of \textbf{U} are orthogonal to \textbf{G}, $V_g$ and $V_u$ are orthogonal subspaces of $\Re^N$. From this fact together with $c$+$f$=$N$ follows now  $V_g \times V_u = \Re^N$. Obviously all normal modes with $\kappa=+1$ ($-1$) must belong to the subspace $V_g$ ($V_u$), simultaneously they must also be a basis to the subspaces as the set of all vectors describing the normal modes must be a basis for the $\Re^N$. The total number of normal modes belonging to one distinct eigenvalue of $\Sigma$ is therefore preserved, independently of the actual choice of the normal modes.}\BibitemShut {Stop}%
\bibitem [{\citenamefont {Plechinger}\ \emph {et~al.}(2012)\citenamefont
  {Plechinger}, \citenamefont {Heydrich}, \citenamefont {Eroms}, \citenamefont
  {Weiss}, \citenamefont {Sch\"{u}ller},\ and\ \citenamefont
  {Korn}}]{Plechinger2012}%
  \BibitemOpen
  \bibfield  {author} {\bibinfo {author} {\bibfnamefont {G.}~\bibnamefont
  {Plechinger}}, \bibinfo {author} {\bibfnamefont {S.}~\bibnamefont
  {Heydrich}}, \bibinfo {author} {\bibfnamefont {J.}~\bibnamefont {Eroms}},
  \bibinfo {author} {\bibfnamefont {D.}~\bibnamefont {Weiss}}, \bibinfo
  {author} {\bibfnamefont {C.}~\bibnamefont {Sch\"{u}ller}}, \ and\ \bibinfo
  {author} {\bibfnamefont {T.}~\bibnamefont {Korn}},\ }\href {\doibase
  10.1063/1.4751266} {\bibfield  {journal} {\bibinfo  {journal} {Applied
  Physics Letters}\ }\textbf {\bibinfo {volume} {101}},\ \bibinfo {pages}
  {101906} (\bibinfo {year} {2012})}\BibitemShut {NoStop}%
\bibitem [{\citenamefont {Zeng}\ \emph {et~al.}(2012)\citenamefont {Zeng},
  \citenamefont {Zhu}, \citenamefont {Liu}, \citenamefont {Fan}, \citenamefont
  {Cui},\ and\ \citenamefont {Zhang}}]{Zeng2012a}%
  \BibitemOpen
  \bibfield  {author} {\bibinfo {author} {\bibfnamefont {H.}~\bibnamefont
  {Zeng}}, \bibinfo {author} {\bibfnamefont {B.}~\bibnamefont {Zhu}}, \bibinfo
  {author} {\bibfnamefont {K.}~\bibnamefont {Liu}}, \bibinfo {author}
  {\bibfnamefont {J.}~\bibnamefont {Fan}}, \bibinfo {author} {\bibfnamefont
  {X.}~\bibnamefont {Cui}}, \ and\ \bibinfo {author} {\bibfnamefont {Q.~M.}\
  \bibnamefont {Zhang}},\ }\href {\doibase 10.1103/PhysRevB.86.241301}
  {\bibfield  {journal} {\bibinfo  {journal} {Physical Review B}\ }\textbf
  {\bibinfo {volume} {86}},\ \bibinfo {pages} {241301} (\bibinfo {year}
  {2012})}\BibitemShut {NoStop}%
\bibitem [{\citenamefont {Zhang}\ \emph {et~al.}(2013)\citenamefont {Zhang},
  \citenamefont {Han}, \citenamefont {Wu}, \citenamefont {Milana},
  \citenamefont {Lu}, \citenamefont {Li}, \citenamefont {Ferrari},\ and\
  \citenamefont {Tan}}]{Zhang2013}%
  \BibitemOpen
  \bibfield  {author} {\bibinfo {author} {\bibfnamefont {X.}~\bibnamefont
  {Zhang}}, \bibinfo {author} {\bibfnamefont {W. P.}~\bibnamefont {Han}}, \bibinfo
  {author} {\bibfnamefont {J. B.}~\bibnamefont {Wu}}, \bibinfo {author}
  {\bibfnamefont {S.}~\bibnamefont {Milana}}, \bibinfo {author} {\bibfnamefont
  {Y.}~\bibnamefont {Lu}}, \bibinfo {author} {\bibfnamefont {Q. Q.}~\bibnamefont
  {Li}}, \bibinfo {author} {\bibfnamefont {A. C.}~\bibnamefont {Ferrari}}, \ and\
  \bibinfo {author} {\bibfnamefont {P. H.}~\bibnamefont {Tan}},\ }\href {\doibase
  10.1103/PhysRevB.87.115413} {\bibfield  {journal} {\bibinfo  {journal}
  {Physical Review B}\ }\textbf {\bibinfo {volume} {87}},\ \bibinfo {pages}
  {115413} (\bibinfo {year} {2013})}\BibitemShut {NoStop}%
\bibitem [{\citenamefont {Malard}\ \emph {et~al.}(2009)\citenamefont {Malard},
  \citenamefont {Guimar\~{a}es}, \citenamefont {Mafra}, \citenamefont
  {Mazzoni},\ and\ \citenamefont {Jorio}}]{Malard2009a}%
  \BibitemOpen
  \bibfield  {author} {\bibinfo {author} {\bibfnamefont {L.~M.}\ \bibnamefont
  {Malard}}, \bibinfo {author} {\bibfnamefont {M.~H.~D.}\ \bibnamefont
  {Guimar\~{a}es}}, \bibinfo {author} {\bibfnamefont {D.~L.}\ \bibnamefont
  {Mafra}}, \bibinfo {author} {\bibfnamefont {M.~S.~C.}\ \bibnamefont
  {Mazzoni}}, \ and\ \bibinfo {author} {\bibfnamefont {A.}~\bibnamefont
  {Jorio}},\ }\href {\doibase 10.1103/PhysRevB.79.125426} {\bibfield  {journal}
  {\bibinfo  {journal} {Physical Review B}\ }\textbf {\bibinfo {volume} {79}},\ \bibinfo {pages} {125426}
  (\bibinfo {year} {2009})}\BibitemShut {NoStop}%
\bibitem [{\citenamefont {Michel}\ and\ \citenamefont
  {Verberck}(2008)}]{Michel2008}%
  \BibitemOpen
  \bibfield  {author} {\bibinfo {author} {\bibfnamefont {K.~H.}\ \bibnamefont
  {Michel}}\ and\ \bibinfo {author} {\bibfnamefont {B.}~\bibnamefont
  {Verberck}},\ }\href {\doibase 10.1103/PhysRevB.78.085424} {\bibfield
  {journal} {\bibinfo  {journal} {Physical Review B}\ }\textbf {\bibinfo {volume} {78}},\ \bibinfo {pages} {085424}
  (\bibinfo {year} {2008})}\BibitemShut {NoStop}%
\bibitem [{\citenamefont {Tan}\ \emph {et~al.}(2012)\citenamefont {Tan},
  \citenamefont {Han}, \citenamefont {Zhao}, \citenamefont {Wu}, \citenamefont
  {Chang}, \citenamefont {Wang}, \citenamefont {Wang}, \citenamefont {Bonini},
  \citenamefont {Marzari}, \citenamefont {Pugno}, \citenamefont {Savini},
  \citenamefont {Lombardo},\ and\ \citenamefont {Ferrari}}]{Tan2012}%
  \BibitemOpen
  \bibfield  {author} {\bibinfo {author} {\bibfnamefont {P.~H.}\ \bibnamefont
  {Tan}}, \bibinfo {author} {\bibfnamefont {W.~P.}\ \bibnamefont {Han}},
  \bibinfo {author} {\bibfnamefont {W.~J.}\ \bibnamefont {Zhao}}, \bibinfo
  {author} {\bibfnamefont {Z.~H.}\ \bibnamefont {Wu}}, \bibinfo {author}
  {\bibfnamefont {K.}~\bibnamefont {Chang}}, \bibinfo {author} {\bibfnamefont
  {H.}~\bibnamefont {Wang}}, \bibinfo {author} {\bibfnamefont {Y.~F.}\
  \bibnamefont {Wang}}, \bibinfo {author} {\bibfnamefont {N.}~\bibnamefont
  {Bonini}}, \bibinfo {author} {\bibfnamefont {N.}~\bibnamefont {Marzari}},
  \bibinfo {author} {\bibfnamefont {N.}~\bibnamefont {Pugno}}, \bibinfo
  {author} {\bibfnamefont {G.}~\bibnamefont {Savini}}, \bibinfo {author}
  {\bibfnamefont {A.}~\bibnamefont {Lombardo}}, \ and\ \bibinfo {author}
  {\bibfnamefont {A.~C.}\ \bibnamefont {Ferrari}},\ }\href {\doibase
  10.1038/nmat3245} {\bibfield  {journal} {\bibinfo  {journal} {Nature
  Materials}\ }\textbf {\bibinfo {volume} {11}},\ \bibinfo {pages} {294}
  (\bibinfo {year} {2012})}\BibitemShut {NoStop}%
\bibitem [{\citenamefont {Herziger}\ \emph {et~al.}(2012)\citenamefont
  {Herziger}, \citenamefont {May},\ and\ \citenamefont
  {Maultzsch}}]{Herziger2012}%
  \BibitemOpen
  \bibfield  {author} {\bibinfo {author} {\bibfnamefont {F.}~\bibnamefont
  {Herziger}}, \bibinfo {author} {\bibfnamefont {P.}~\bibnamefont {May}}, \
  and\ \bibinfo {author} {\bibfnamefont {J.}~\bibnamefont {Maultzsch}},\ }\href
  {\doibase 10.1103/PhysRevB.85.235447} {\bibfield  {journal} {\bibinfo
  {journal} {Physical Review B}\ }\textbf {\bibinfo {volume} {85}},\ \bibinfo
  {pages} {235447} (\bibinfo {year} {2012})}\BibitemShut {NoStop}%
\bibitem [{\citenamefont {Lui}\ and\ \citenamefont {Heinz}(2013)}]{Lui2013}%
  \BibitemOpen
  \bibfield  {author} {\bibinfo {author} {\bibfnamefont {C.~H.}\ \bibnamefont
  {Lui}}\ and\ \bibinfo {author} {\bibfnamefont {T.~F.}\ \bibnamefont
  {Heinz}},\ }\href {\doibase 10.1103/PhysRevB.87.121404} {\bibfield  {journal}
  {\bibinfo  {journal} {Physical Review B}\ }\textbf {\bibinfo {volume} {87}},\
  \bibinfo {pages} {121404} (\bibinfo {year} {2013})}\BibitemShut {NoStop}%
\bibitem [{\citenamefont {Luo}\ \emph {et~al.}(1996)\citenamefont {Luo},
  \citenamefont {Ruggerone},\ and\ \citenamefont {Toennies}}]{Luo1996}%
  \BibitemOpen
  \bibfield  {author} {\bibinfo {author} {\bibfnamefont {N. S.}~\bibnamefont
  {Luo}}, \bibinfo {author} {\bibfnamefont {P.}~\bibnamefont {Ruggerone}}, \
  and\ \bibinfo {author} {\bibfnamefont {J. P.}~\bibnamefont {Toennies}},\ }\href
  {\doibase 10.1103/PhysRevB.54.5051} {\bibfield  {journal} {\bibinfo
  {journal} {Physical Review B}\ }\textbf {\bibinfo {volume} {54}},\ \bibinfo
  {pages} {5051} (\bibinfo {year} {1996})}\BibitemShut {NoStop}%
\bibitem [{\citenamefont {Gillen}\ \emph {et~al.}(2009)\citenamefont {Gillen},
  \citenamefont {Mohr}, \citenamefont {Thomsen},\ and\ \citenamefont
  {Maultzsch}}]{Gillen2009}%
  \BibitemOpen
  \bibfield  {author} {\bibinfo {author} {\bibfnamefont {R.}~\bibnamefont
  {Gillen}}, \bibinfo {author} {\bibfnamefont {M.}~\bibnamefont {Mohr}},
  \bibinfo {author} {\bibfnamefont {C.}~\bibnamefont {Thomsen}}, \ and\
  \bibinfo {author} {\bibfnamefont {J.}~\bibnamefont {Maultzsch}},\ }\href
  {\doibase 10.1103/PhysRevB.80.155418} {\bibfield  {journal} {\bibinfo
  {journal} {Physical Review B}\ }\textbf {\bibinfo {volume} {80}},\ \bibinfo
  {pages} {155418} (\bibinfo {year} {2009})}\BibitemShut {NoStop}%
\bibitem [{Note2()}]{Note2}%
  \BibitemOpen
  \bibinfo {note} {Control calculations for slabs of 1-5 layers of MoS$_2$ and
  WS$_2$ were performed in the frame of density functional (perturbation)
  theory on the level of the PBEsol exchange-correlation functional as
  implemented into the Quantum Espresso suite. The electrostatic potential
  between valence electrons and chemically inert atomic core of Mo, W and S was
  modeled by the recently proposed GBRV high-throughput pseudopotentials\cite
  {Garrity2014} with a cutoff of 80 Ry. All reciprocal space integrations were
  performed by a discrete k-point sampling of 24x24x1 k-points in the Brillouin
  zone. We fully optimized the atomic positions and cell parameters of the
  considered systems until the residual forces between atoms was smaller than
  0.001 eV/\r A and the pressure on the cell was lower than 0.001 GPa.
  Interactions of the slabs with residual periodic images due to the 3D
  boundary conditions were minimized by maintaining a vacuum layer between
  slabs of at least 25\r A. After obtaining a converged electronic ground
  state, the frequencies and atomic displacements of the $\Gamma $-point
  phonons was calculated through density functional perturbation theory
  (DFPT).}\BibitemShut {Stop}%
	
\bibitem [{\citenamefont {Staiger}\ \emph {et~al.}(2015)\citenamefont
  {Staiger}, \citenamefont {Gillen}, \citenamefont {Scheuschner}, \citenamefont
  {Ochedowski}, \citenamefont {Schleberger}, \citenamefont {Thomsen},\ and\
  \citenamefont {Maultzsch}}]{Staiger2015}%
  \BibitemOpen
  \bibfield  {author} {\bibinfo {author} {\bibfnamefont {M.}~\bibnamefont
  {Staiger}}, \bibinfo {author} {\bibfnamefont {R.}~\bibnamefont {Gillen}},
  \bibinfo {author} {\bibfnamefont {N.}~\bibnamefont {Scheuschner}}, \bibinfo
  {author} {\bibfnamefont {O.}~\bibnamefont {Ochedowski}}, \bibinfo {author}
  {\bibfnamefont {M.}~\bibnamefont {Schleberger}}, \bibinfo {author}
  {\bibfnamefont {C.}~\bibnamefont {Thomsen}}, \ and\ \bibinfo {author}
  {\bibfnamefont {J.}~\bibnamefont {Maultzsch}},\ } {\bibfield  {journal} {\bibinfo
  {journal} {submitted to Physical Review B}} {\  (\bibinfo  {year} {2015})}\BibitemShut {Stop}%


\bibitem [{\citenamefont {Scheuschner}\ \emph {et~al.}(2012)\citenamefont
  {Scheuschner}, \citenamefont {Ochedowski}, \citenamefont {Schleberger},\ and\
  \citenamefont {Maultzsch}}]{Scheuschner2012a}%
  \BibitemOpen
  \bibfield  {author} {\bibinfo {author} {\bibfnamefont {N.}~\bibnamefont
  {Scheuschner}}, \bibinfo {author} {\bibfnamefont {O.}~\bibnamefont
  {Ochedowski}}, \bibinfo {author} {\bibfnamefont {M.}~\bibnamefont
  {Schleberger}}, \ and\ \bibinfo {author} {\bibfnamefont {J.}~\bibnamefont
  {Maultzsch}},\ }\href {\doibase 10.1002/pssb.201200389} {\bibfield  {journal}
  {\bibinfo  {journal} {Physica Status Solidi (B)}\ }\textbf {\bibinfo {volume}
  {249}},\ \bibinfo {pages} {2644} (\bibinfo {year} {2012})}\BibitemShut
  {NoStop}%
\bibitem [{\citenamefont {Dhakal}\ \emph {et~al.}(2014)\citenamefont {Dhakal},
  \citenamefont {Duong}, \citenamefont {Lee}, \citenamefont {Nam},
  \citenamefont {Kim}, \citenamefont {Kan}, \citenamefont {Lee},\ and\
  \citenamefont {Kim}}]{Dhakal2014}%
  \BibitemOpen
  \bibfield  {author} {\bibinfo {author} {\bibfnamefont {K.~P.}\ \bibnamefont
  {Dhakal}}, \bibinfo {author} {\bibfnamefont {D.~L.}\ \bibnamefont {Duong}},
  \bibinfo {author} {\bibfnamefont {J.}~\bibnamefont {Lee}}, \bibinfo {author}
  {\bibfnamefont {H.}~\bibnamefont {Nam}}, \bibinfo {author} {\bibfnamefont
  {M.}~\bibnamefont {Kim}}, \bibinfo {author} {\bibfnamefont {M.}~\bibnamefont
  {Kan}}, \bibinfo {author} {\bibfnamefont {Y.~H.}\ \bibnamefont {Lee}}, \ and\
  \bibinfo {author} {\bibfnamefont {J.}~\bibnamefont {Kim}},\ }\href {\doibase
  10.1039/c4nr03703k} {\bibfield  {journal} {\bibinfo  {journal} {Nanoscale}\
  }\textbf {\bibinfo {volume} {6}},\ \bibinfo {pages} {13028} (\bibinfo {year}
  {2014})}\BibitemShut {NoStop}%
\bibitem [{\citenamefont {Livneh}\ and\ \citenamefont
  {Spanier}(2014)}]{Livneh2014}%
  \BibitemOpen
  \bibfield  {author} {\bibinfo {author} {\bibfnamefont {T.}~\bibnamefont
  {Livneh}}\ and\ \bibinfo {author} {\bibfnamefont {J.}~\bibnamefont
  {Spanier}},\ }\href {http://arxiv.org/abs/1408.6748} {\bibfield  {journal}
  {\bibinfo  {journal} {arXiv preprint}\ } (\bibinfo {year}
  {2014})},\ \Eprint {http://arxiv.org/abs/1408.6748} {arXiv:1408.6748}
  \BibitemShut {NoStop}%
\bibitem [{\citenamefont {Molina-S\'{a}nchez}\ \emph
  {et~al.}(2013)\citenamefont {Molina-S\'{a}nchez}, \citenamefont {Sangalli},
  \citenamefont {Hummer}, \citenamefont {Marini},\ and\ \citenamefont
  {Wirtz}}]{Molina-Sanchez2013}%
  \BibitemOpen
  \bibfield  {author} {\bibinfo {author} {\bibfnamefont {A.}~\bibnamefont
  {Molina-S\'{a}nchez}}, \bibinfo {author} {\bibfnamefont {D.}~\bibnamefont
  {Sangalli}}, \bibinfo {author} {\bibfnamefont {K.}~\bibnamefont {Hummer}},
  \bibinfo {author} {\bibfnamefont {A.}~\bibnamefont {Marini}}, \ and\ \bibinfo
  {author} {\bibfnamefont {L.}~\bibnamefont {Wirtz}},\ }\href {\doibase
  10.1103/PhysRevB.88.045412} {\bibfield  {journal} {\bibinfo  {journal}
  {Physical Review B}\ }\textbf {\bibinfo {volume} {88}},\ \bibinfo {pages} {045412}
  (\bibinfo {year} {2013})}\BibitemShut {NoStop}%
\bibitem [{\citenamefont {Schuller}\ \emph {et~al.}(2013)\citenamefont
  {Schuller}, \citenamefont {Karaveli}, \citenamefont {Schiros}, \citenamefont
  {He}, \citenamefont {Yang}, \citenamefont {Kymissis}, \citenamefont {Shan},\
  and\ \citenamefont {Zia}}]{Schuller2013}%
  \BibitemOpen
  \bibfield  {author} {\bibinfo {author} {\bibfnamefont {J.~A.}\ \bibnamefont
  {Schuller}}, \bibinfo {author} {\bibfnamefont {S.}~\bibnamefont {Karaveli}},
  \bibinfo {author} {\bibfnamefont {T.}~\bibnamefont {Schiros}}, \bibinfo
  {author} {\bibfnamefont {K.}~\bibnamefont {He}}, \bibinfo {author}
  {\bibfnamefont {S.}~\bibnamefont {Yang}}, \bibinfo {author} {\bibfnamefont
  {I.}~\bibnamefont {Kymissis}}, \bibinfo {author} {\bibfnamefont
  {J.}~\bibnamefont {Shan}}, \ and\ \bibinfo {author} {\bibfnamefont
  {R.}~\bibnamefont {Zia}},\ }\href {\doibase 10.1038/nnano.2013.20} {\bibfield
   {journal} {\bibinfo  {journal} {Nature Nanotechnology}\ }\textbf {\bibinfo
  {volume} {8}},\ \bibinfo {pages} {271} (\bibinfo {year} {2013})}\BibitemShut
  {NoStop}%
\bibitem [{\citenamefont {Bradley}\ \emph {et~al.}(2015)\citenamefont
  {Bradley}, \citenamefont {{M. Ugeda}}, \citenamefont {da~Jornada},
  \citenamefont {Qiu}, \citenamefont {Ruan}, \citenamefont {Zhang},
  \citenamefont {Wickenburg}, \citenamefont {Riss}, \citenamefont {Lu},
  \citenamefont {Mo}, \citenamefont {Hussain}, \citenamefont {Shen},
  \citenamefont {Louie},\ and\ \citenamefont {Crommie}}]{Bradley2015}%
  \BibitemOpen
  \bibfield  {author} {\bibinfo {author} {\bibfnamefont {A.~J.}\ \bibnamefont
  {Bradley}}, \bibinfo {author} {\bibfnamefont {M.}~\bibnamefont {{M. Ugeda}}},
  \bibinfo {author} {\bibfnamefont {F.~H.}\ \bibnamefont {da~Jornada}},
  \bibinfo {author} {\bibfnamefont {D.~Y.}\ \bibnamefont {Qiu}}, \bibinfo
  {author} {\bibfnamefont {W.}~\bibnamefont {Ruan}}, \bibinfo {author}
  {\bibfnamefont {Y.}~\bibnamefont {Zhang}}, \bibinfo {author} {\bibfnamefont
  {S.}~\bibnamefont {Wickenburg}}, \bibinfo {author} {\bibfnamefont
  {A.}~\bibnamefont {Riss}}, \bibinfo {author} {\bibfnamefont {J.}~\bibnamefont
  {Lu}}, \bibinfo {author} {\bibfnamefont {S.-K.}\ \bibnamefont {Mo}}, \bibinfo
  {author} {\bibfnamefont {Z.}~\bibnamefont {Hussain}}, \bibinfo {author}
  {\bibfnamefont {Z.-X.}\ \bibnamefont {Shen}}, \bibinfo {author}
  {\bibfnamefont {S.~G.}\ \bibnamefont {Louie}}, \ and\ \bibinfo {author}
  {\bibfnamefont {M.~F.}\ \bibnamefont {Crommie}},\ }\href {\doibase
  10.1021/acs.nanolett.5b00160} {\bibfield  {journal} {\bibinfo  {journal}
  {Nano Letters}\ } (\bibinfo {year} {2015}),\
  10.1021/acs.nanolett.5b00160}\BibitemShut {NoStop}%
\bibitem [{\citenamefont {Bergh\"{a}user}\ and\ \citenamefont
  {Malic}(2014)}]{Berghauser2014}%
  \BibitemOpen
  \bibfield  {author} {\bibinfo {author} {\bibfnamefont {G.}~\bibnamefont
  {Bergh\"{a}user}}\ and\ \bibinfo {author} {\bibfnamefont {E.}~\bibnamefont
  {Malic}},\ }\href {\doibase 10.1103/PhysRevB.89.125309} {\bibfield  {journal}
  {\bibinfo  {journal} {Physical Review B}\ }\textbf {\bibinfo {volume} {89}},\
  \bibinfo {pages} {125309} (\bibinfo {year} {2014})}\BibitemShut {NoStop}%
\bibitem [{\citenamefont {Lee}\ \emph {et~al.}(2010{\natexlab{b}})\citenamefont
  {Lee}, \citenamefont {Yan}, \citenamefont {Brus}, \citenamefont {Heinz},
  \citenamefont {Hone},\ and\ \citenamefont {Ryu}}]{Lee2010}%
  \BibitemOpen
  \bibfield  {author} {\bibinfo {author} {\bibfnamefont {C.}~\bibnamefont
  {Lee}}, \bibinfo {author} {\bibfnamefont {H.}~\bibnamefont {Yan}}, \bibinfo
  {author} {\bibfnamefont {L.~E.}\ \bibnamefont {Brus}}, \bibinfo {author}
  {\bibfnamefont {T.~F.}\ \bibnamefont {Heinz}}, \bibinfo {author}
  {\bibfnamefont {J.}~\bibnamefont {Hone}}, \ and\ \bibinfo {author}
  {\bibfnamefont {S.}~\bibnamefont {Ryu}},\ }\href {\doibase 10.1021/nn1003937}
  {\bibfield  {journal} {\bibinfo  {journal} {ACS Nano}\ }\textbf {\bibinfo
  {volume} {4}},\ \bibinfo {pages} {2695} (\bibinfo {year}
  {2010}{\natexlab{b}})}\BibitemShut {NoStop}%
\bibitem [{\citenamefont {Ruppert}\ \emph {et~al.}(2014)\citenamefont
  {Ruppert}, \citenamefont {Aslan},\ and\ \citenamefont {Heinz}}]{Ruppert2014}%
  \BibitemOpen
  \bibfield  {author} {\bibinfo {author} {\bibfnamefont {C.}~\bibnamefont
  {Ruppert}}, \bibinfo {author} {\bibfnamefont {O.~B.}\ \bibnamefont {Aslan}},
  \ and\ \bibinfo {author} {\bibfnamefont {T.~F.}\ \bibnamefont {Heinz}},\
  }\href@noop {} {\bibfield  {journal} {\bibinfo  {journal} {Nano Letters}\
  }\textbf {\bibinfo {volume} {14}},\ \bibinfo {pages} {6231} (\bibinfo {year}
  {2014})}\BibitemShut {NoStop}%
\bibitem [{\citenamefont {Garrity}\ \emph {et~al.}(2014)\citenamefont
  {Garrity}, \citenamefont {Bennett}, \citenamefont {Rabe},\ and\ \citenamefont
  {Vanderbilt}}]{Garrity2014}%
  \BibitemOpen
  \bibfield  {author} {\bibinfo {author} {\bibfnamefont {K.~F.}\ \bibnamefont
  {Garrity}}, \bibinfo {author} {\bibfnamefont {J.~W.}\ \bibnamefont
  {Bennett}}, \bibinfo {author} {\bibfnamefont {K.~M.}\ \bibnamefont {Rabe}}, \
  and\ \bibinfo {author} {\bibfnamefont {D.}~\bibnamefont {Vanderbilt}},\
  }\href {\doibase 10.1016/j.commatsci.2013.08.053} {\bibfield  {journal}
  {\bibinfo  {journal} {Computational Materials Science}\ }\textbf {\bibinfo
  {volume} {81}},\ \bibinfo {pages} {446} (\bibinfo {year} {2014})} \BibitemShut {NoStop}}%
\end{thebibliography}
\end{document}